\documentclass[pra,letterpaper,twocolumn,aps,footinbib,superscriptaddress,showpacs,babel]{revtex4-1}

\usepackage{amsmath, upgreek, amssymb, graphicx, subfigure, paralist, dsfont, transparent}
\usepackage[colorlinks, linkcolor=blue, citecolor=blue, urlcolor=blue, breaklinks=true]{hyperref}
\usepackage{xspace}

\newcommand{\ket}[1]{|#1\rangle}
\newcommand{\bra}[1]{\langle #1|}

\newcommand{\sm}{\sigma^-}
\newcommand{\spl}{\sigma^+}
\newcommand{\sx}{\sigma_x}
\newcommand{\sy}{\sigma_y}
\newcommand{\sz}{\sigma_z}

\newcommand{\dc}{\alpha_i} 		
\newcommand{\dx}{\beta_{ij}}	

\newcommand{\ttime}{t_s}				
\newcommand{\gendrift}{d}			
\newcommand{\gendiff}{D}			
\newcommand{\oneddrift}{\alpha}		
\newcommand{\oneddiff}{D}	

\newcommand{\LD}{\eta}				
\newcommand{\Nph}{N_{ph}}				
\newcommand{\pin}{P_0}				
\newcommand{\wres}{\omega_0}		
\newcommand{\wlaser}{\omega_L}	
\newcommand{\wmode}{\nu}				
\newcommand{\tpulse}{t_p}				
\newcommand{\tspec}{\tau} 
\newcommand{\tr}[1]{\mathrm{Tr} \left[ #1 \right]}  
\newcommand{\para}{\Delta_0}

\begin{document}

\title{Photon Recoil Spectroscopy: Systematic Shifts and Nonclassical Enhancements}

\author{M. Schulte}
\affiliation{Institute for Theoretical Physics and Institute for Gravitational Physics (Albert-Einstein-Institute), Leibniz University Hannover, Appelstrasse 2, 30167 Hannover, Germany}

\author{N. L{\"o}rch}
\affiliation{Department of Physics, University of Basel, Klingelbergstrasse 82, CH-4056 Basel, Switzerland}


\author{P. O. Schmidt}
\affiliation{Physikalisch-Technische Bundesanstalt, 38116 Braunschweig, Germany}
\affiliation{Institute for Quantum Optics, Leibniz University Hannover, Welfengarten 1, 30167 Hannover, Germany}

\author{K. Hammerer}
\affiliation{Institute for Theoretical Physics and Institute for Gravitational Physics (Albert-Einstein-Institute), Leibniz University Hannover, Appelstrasse 2, 30167 Hannover, Germany}

\begin{abstract}
In photon recoil spectroscopy, signals are extracted from recoils imparted by the spectroscopy light on the motion of trapped ions as demonstrated by C. Hempel et al., Nature Photonics {\bf 7}, 630 (2013) and Y. Wan et al., Nature Communications {\bf 5} 3096 (2014). The method exploits the exquisite efficiency in the detection of phonons achievable in ion crystals, and is thus particularly suitable for species with broad non-cycling transitions where detection of fluorescence photons is impractical. Here, we develop a theoretical model for the description of photon recoil spectroscopy based on a Fokker-Planck equation for the Wigner function of the phonon mode.  Our model correctly explains systematic shifts due to Doppler heating and cooling as observed in the experiment. Furthermore, we investigate quantum metrological schemes for enhancing the spectroscopic sensitivity based on the preparation and detection of nonclassical states of the phonon mode.

%
\end{abstract}

\date\today

\maketitle

\section{Introduction}\label{sec:Intro}
The remarkable accuracy of precision spectroscopy with well-controlled quantum systems opens up possibilities to investigate fundamental questions in physics\,\cite{SafronovaEtAl2018}.
In particular molecules and molecular ions enable to precisely test the validity of theories like quantum electrodynamics\,\cite{KoelemeijEtAl2007, HerrmannEtAl2009, AsvanyEtAl2008}, search for a possible temporal variation of fundamental constants\,\cite{SchillerKorobov2005, BeloyEtAl2011, KajitaEtAl2014} or an electrical dipole moment of the electron\,\cite{CairncrossEtAl2017, LohEtAl2013}, and test parity violations\,\cite{BorschevskyEtAl2012}.
For this, single trapped ions form an ideal platform in which external distortions are minimal and systematic errors can be reduced by a high degree of control over internal and external degrees of freedom\,\cite{Wineland2013, BlattWineland2008, LeibfriedBlattMonroeEtAl2003}.
By combining it with a second, well-controlled ion species, even ions with a complex electronic level structure such as molecular ions can be investigated.
Using variations of the quantum logic approach, first implemented in the aluminium optical clock\,\cite{Schmidt2005, RosenbandEtAl2008}, enables the sympathetic preparation of electronic and motional states\,\cite{KielpinskiKingMyattEtAl2000, WanGebertWolfEtAl2015, ChouEtAl2017} as well as efficient state detection\,\cite{GebertWanWolfEtAl2016, WolfWanHeipEtAl2016} and high-resolution spectroscopy\,\cite{WanGebertWuebbenaEtAl2014, ChouEtAl2017}.
In particular, frequency measurement techniques based on the recoil of absorbed photons have recently been demonstrated\,\cite{HempelLanyonJurcevicEtAl2013, WanGebertWuebbenaEtAl2014, ClosEtAl2014}.
Important for the use in precision spectroscopy is good knowledge on any systematic shifts induced by the detection scheme and additionally, high sensitivity is desirable to detect small signals of only a few absorbed photons as in the case of open transitions in molecular ions.

Therefore, we develop in chapter \ref{sec:TheoreticalModel} a theoretical model of photon recoil spectroscopy (PRS), and in particular of the scheme implemented recently by Wan et al. \cite{WanGebertWuebbenaEtAl2014}. We use phase space models to take into account drifts and diffusion of the quantum mechanical motion of the ion during interaction with spectroscopy light. The photon recoils will set the ion in motion according to a Fokker-Planck equation with drift and diffusion coefficients which both exhibit a nonlinear dependence on the ion's momentum (due to the Doppler effect) and on the detuning of the spectroscopy laser. Thus both, drift and diffusion, contribute to the spectroscopic signal. Based on the Fokker-Planck equation we explain in chapter \ref{sec:Shift} how an observed systematic shift of the measured resonance frequency results as a subtle consequence of the Doppler effect. We examine in chapter \ref{sec:GainInSpec} quantum metrological schemes where nonclassical states of motion are used to enhance the spectroscopic sensitivity. We discuss the gain achievable with various nonclassical motional states, as was demonstrated by Hempel et al.\,\cite{HempelLanyonJurcevicEtAl2013}, such as squeezed states, cat states or Fock states, taking into account nonlinear drift and diffusion in phase space. We determine the Fisher Information implied by the Fokker-Planck equation, and infer the intrinsic limitations of the nonclassical enhancement due to the diffusive nature of the ion's motion. One main conclusion is that all nonclassical states considered here provide a similar enhancement in sensitivity for small average phonon numbers (below about 10) of the initial state. For large phonon numbers squeezed states or Fock states are much more robust to diffusive noise than cat states and can still give a significant enhancement. We use these conclusions to discuss the degree of squeezing required for spectroscopy based on recoils of a \emph{single} photon. Details of calculations are collected in Appendix \ref{sec:app-OBE} to \ref{sec:app-average}.

\section{Spectroscopy via photon recoils} \label{sec:TheoreticalModel}

In this section we introduce PRS and show that the dynamics can be described as a combination of drift and diffusion, that is, displacements and broadenings of the quasi-probability distribution in the phase space of the ion's motional degree of freedom.
First we develop a theoretical model for the laser-ion interactions during spectroscopy as initially described in\,\cite{LoerchThesis2015}.
A main result of this section is that the evolution of an initial motional state is governed by a Fokker-Planck equation for the Wigner function and the derivation of the drift and diffusion coefficients from the spectroscopy interaction.

\subsection{Concept of PRS}
In a typical experimental setting for PRS, two ions of different species are confined in a linear Paul trap. The electronic transition frequency of one of the two ions (referred to as the spectroscopy ion) is to be determined. The other ion (referred to as logic ion) has a known cycling transition for laser cooling and long-lived internal states which can be manipulated using e.g. laser radiation. By joint manipulation of internal and motional states, different motional states can be prepared and read-out. In particular, the logic ion can be used as a phonon detector. As a consequence of the strong Coulomb interaction between them, the motion of the ions can be treated as two coupled three-dimensional harmonic oscillators\,\cite{LeibfriedBlattMonroeEtAl2003}. In a linear ion trap with strong radial confinement, the ions will arrange in a linear string along the trap axis and perform small oscillations around their equilibrium positions that decouple into axial and radial modes\,\cite{MorigiWalther2001, KielpinskiKingMyattEtAl2000}.
Assuming that the spectroscopy laser is directed along the axial direction, the description is reduced to a one-dimensional model of the motion in this direction. We call the two modes along this direction the in- and out-of-phase modes which can be distinguished unambiguously by their oscillation frequency. Therefore we reduce our model to the description of a single phonon mode. However, this mode refers to the collective motion of the two-species, two-ion crystal with different mode amplitudes of the two ions that depend on the mass ratio\,\cite{KielpinskiKingMyattEtAl2000, MorigiWalther2001, WuebbenaEtAl2012}.


\begin{figure}[tb]
\centering
\includegraphics[width=\linewidth]{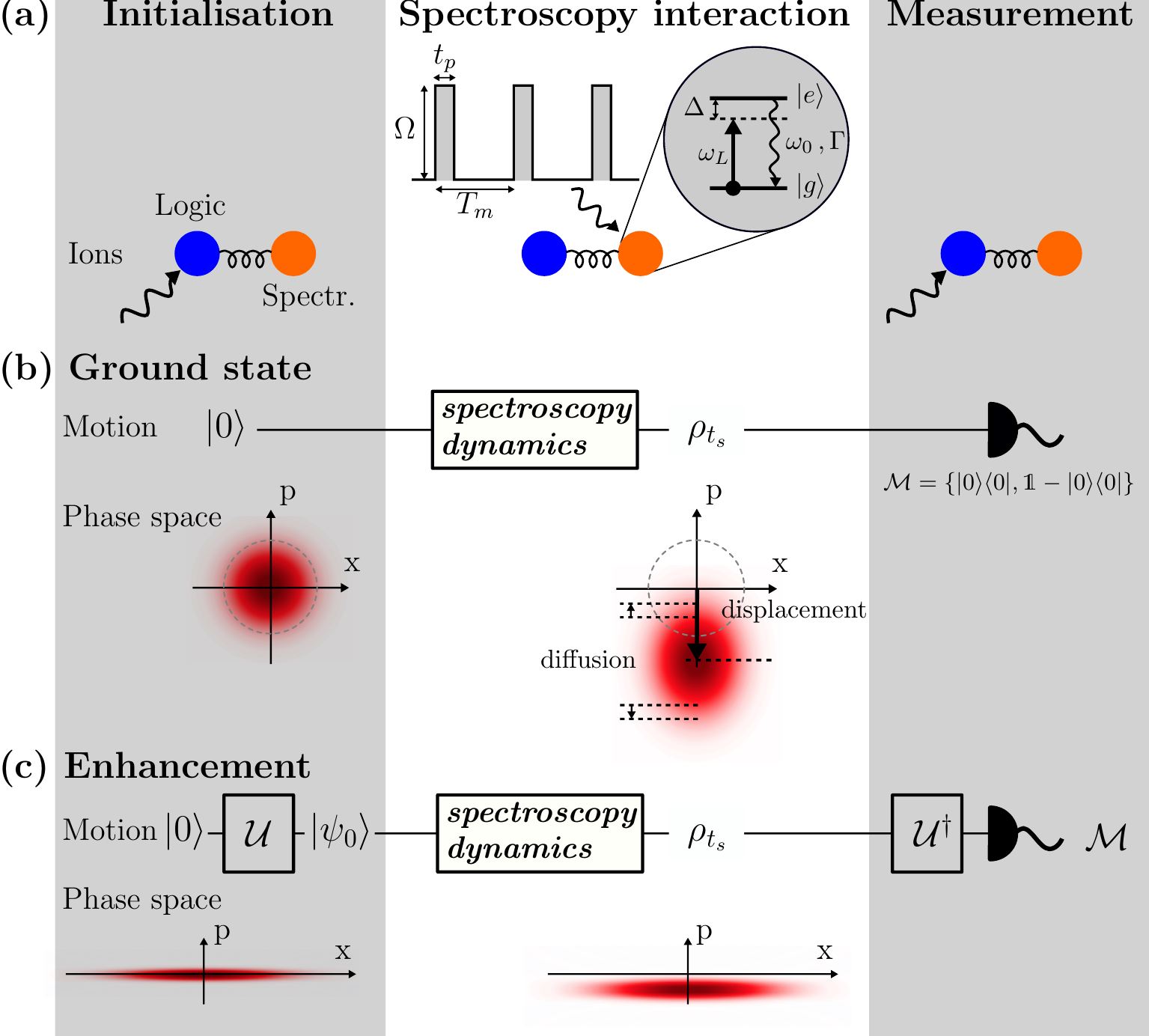}
\caption{(a) Schematic overview of PRS consisting of initial state preparation on the logic ion, pulsed laser-ion interaction during spectroscopy and measurement $\mathcal{M}$ (from left to right column). Part (b) shows spectroscopy with the ground state of motion. The spectroscopy dynamics gives a coherent displacement and incoherent diffusion of the Wigner function in phase space. The measured signal is the overlap $ \bra{0} \rho_{\ttime} \ket{0} $. Enhancing the recoil sensitivity for individual photons is considered in (c). The schematic is for a momentum squeezed state where the overlap with the initial state changes faster under the spectroscopy dynamics.}
\label{fig:scheme}
\end{figure}

Such a composite design allows to overcome limitations of the spectroscopy ion through exploiting favourable properties of the co-trapped logic ion. Notable examples for such a strategy are in state preparation, e.\,g. through sympathetic ground state cooling of different species\,\cite{barrett_sympathetic_2003, home_memory_2009, rugango_sympathetic_2015, wan_efficient_2015}, or in the detection via a state transfer as in quantum logic spectroscopy\,\cite{Schmidt2005}.
In PRS the logic ion will be used for both, initial preparation and final detection of the quantum state of the relevant phonon mode.

A typical measurement sequence of PRS is depicted in Fig.\,\ref{fig:scheme}a. At first, the shared quantized motion is initialized by manipulating the logic ion. The spectroscopy interaction is performed by a series of laser pulses with an (unknown) detuning $\Delta$ from the probed transition of the spectroscopy ion. The pulse duration $\tpulse \ll T_m $ is short compared to the oscillation period $ T_m $ of a selected common mode of the two ions. Each pulse may scatter photons on the spectroscopy transition and consequently transfer momentum in the direction of the laser to the collective motion. The repetition rate of the spectroscopy laser coincides with the normal mode frequency $ \wmode $, i.e. sequential pulses are separated by the oscillation period $T_m = 2\pi/\wmode$ of the two ions. The photon recoil imparted during the spectroscopy dynamics thus adds up in this particular mode while averaging out for other modes. Over the course of several pulses the interaction describes a driving force on the oscillating two ion crystal causing a drift in the amplitude of their common motion. On top of that this force will have fluctuations due to the unavoidable quantum fluctuations of the atomic dipole moment, which will result in a concurrent diffusive motion. In the final step, the logic ion is used again to measure the magnitude of the amplitude of the mode. Finally, by repeating the whole sequence with different detunings $\Delta$, the resonance profile of the spectroscopy ion can be mapped out. From this, the central frequency can be determined via a two point sampling method\,(see Sec.\,\ref{subsec:Measurement}).

On a more conceptual level, depicted in Fig.\,\ref{fig:scheme}b, the procedure is equivalent to the preparation of the common motion in a suitable (pure) initial state $\rho_0$ via the logic ion. E.g. this could be the ground state in the harmonic potential $\rho_0=\ket{0}\bra{0}$  as was the case in the experiment by Wan et al. \cite{WanGebertWuebbenaEtAl2014}. After a certain number of pulses, corresponding to an effective spectroscopy time $\ttime$, the spectroscopy dynamics maps the initial state to a final state $ \rho_{\ttime}$, which will be a displaced and broadened version of the initial state. As the last step of the PRS sequence the overlap $ P = \tr{\rho_0 \rho_{\ttime}} $ between initial state $\rho_0$ and displaced state $ \rho_{\ttime} $ is measured. The spectroscopic signal is then attained from the detuning dependence of this probability $P(\Delta)$.

The idea of a possible enhancement in the spectroscopic signal using squeezed states is shown in Fig.\,\ref{fig:scheme}c. An initial state $\rho_0$ with reduced uncertainty in momentum will exhibit a larger sensitivity to the minute momentum transfer in PRS. Later on we will also consider other states, such as Fock or cat states, and compare their performance under diffusive motion.

\subsection{Evolution during Spectroscopy Dynamics}

The spectroscopy ion is idealized as a two-level system with ground state $\ket{g}$ and excited state $\ket{e}$ for which $\spl$ and $\sm$ are the raising and lowering operators based on the Pauli matrices $\sx, \sy, \sz$.
The Hamiltonian for laser-ion interaction in the dipole approximation and a rotating reference frame with the laser frequency $ \wlaser $ is
\begin{equation}
\label{eq:Hamiltonian} H= \wmode \, a^{\dagger} a - \dfrac{\Delta}{2} \sz + \dfrac{\Omega(t)}{2} \left( \spl e^{i \eta (a + a^{\dagger})} + \mathrm{h.c.} \right)
\end{equation}
Here, $a$ and $a^{\dagger}$ are annihilation and creation operators of the selected normal mode with frequency $ \wmode $ and Lamb-Dicke parameter $\eta$. All other modes are neglected for the reasons explained above.
The Rabi frequency is denoted by $\Omega$ and $ \Delta = \wlaser - \wres $ gives the laser detuning from resonance.
The first two terms in Eq.\,\eqref{eq:Hamiltonian} describe the free evolution of motion and internal states, whereas the last term (interaction term) is only active when the laser pulses are applied.
The pulsed interaction is modelled by a time dependent Rabi frequency $\Omega (t)$ with a box shape as sketched in Fig.\,\ref{fig:scheme}a.
During a pulse of duration $\tpulse$, the density matrix evolves according to the master equation
\begin{equation}
\label{eq:MasterEq} \dot{\rho} = -i \left[H ,\rho \right] + \dfrac{\Gamma}{2} \mathcal{D}\left[ \sm \right] \rho
\end{equation}
with the transition linewidth $\Gamma$ and Lindblad term $ \mathcal{D}\left[ \sm \right] \rho = 2 \sm \rho \spl - \spl \sm \rho - \rho \spl \sm$ describing spontaneous decay to the ground state.

As stated above, to synchronize the laser interactions with the oscillations of the ions and thereby add up the momentum kicks of successive absorptions, each pulse of duration $ \tpulse $ is followed by a free evolution, with $\Omega = 0$, of duration $ T_m - \tpulse $.
Studying the dynamics of Eq.\,\eqref{eq:MasterEq} is now done by introducing a Wigner representation
\begin{equation}
w(x,p) = \dfrac{1}{\pi} \int \mathrm{d}y\, \bra{x+y} \rho \ket{x-y} e^{-2 i y p}
\end{equation}
which is a density matrix for the internal states, depending on the oscillators phase space coordinates $x$ and $p$.
Reducing this representation to either subsystem gives the density matrix
\begin{equation}
\rho_{\mathrm{int}} = \int \mathrm{d}x\, \mathrm{d}p\,\, w(x,p)
\end{equation}
and Wigner function
\begin{equation}\label{eq:Wigner}
W(x,p) = \tr{ w(x,p) }
\end{equation}
respectively.
Equations of motion for $w(x,p)$ are obtained by applying the replacement rules\,\citep{CrispinGardiner2004}
\begin{equation}
\begin{aligned}
\hat{x} \rho &\rightarrow \left(x + \dfrac{i}{2} \dfrac{\partial}{\partial p} \right) w \hspace{0.75cm} \rho \hat{x} \rightarrow \left(x - \dfrac{i}{2} \dfrac{\partial}{\partial p} \right) w \\
\hat{p} \rho &\rightarrow \left(p - \dfrac{i}{2} \dfrac{\partial}{\partial x} \right) w \hspace{0.75cm} \rho \hat{p} \rightarrow \left(p + \dfrac{i}{2} \dfrac{\partial}{\partial x} \right) w
\end{aligned}
\end{equation}
for the quadrature operators.
The interaction terms will then be approximated by the replacements
\begin{equation}
\begin{aligned}
e^{\pm i \eta (a + a^{\dagger}) } \rho &\rightarrow e^{\pm i \bar{\eta} x } \left( 1 \mp \dfrac{\bar{\eta}}{2} \partial_p \right) w\\
\rho e^{\pm i \eta (a + a^{\dagger}) } &\rightarrow e^{\pm i \bar{\eta} x } \left( 1 \pm \dfrac{\bar{\eta}}{2} \partial_p \right) w
\end{aligned}
\end{equation}
where $ \bar{\eta} = \sqrt{2} \eta $ is a re-scaled Lamb-Dicke parameter.
This approach includes terms up to first order in $ \bar{\eta} \partial_p $ and all terms in $\bar{\eta} x$. It is therefore a weaker approximation than the standard Lamb-Dicke approximation, where only first order terms are kept.
With this level of approximation systematic shifts in the spectroscopic signal as observed in\,\cite{WanGebertWuebbenaEtAl2014} can be explained.
The approximate equation of motion is
\begin{equation}
\dot{w} = ( \mathcal{L}_m + \mathcal{L}_a + \mathcal{L}_i) w
\end{equation}
with
\begin{align}
&\mathcal{L}_m w = \wmode (\partial_x p - \partial_p x)\, w \label{eq:dynmot}\\
&\mathcal{L}_a w = -i \left[ -\dfrac{\Delta}{2} \sz + \dfrac{\Omega(t)}{2} \left(\spl e^{i \bar{\eta} x} + h.c. \right), w \right] \\
&\qquad \quad+ \dfrac{\Gamma}{2} \mathcal{D}\left[ \sm \right] w \nonumber \\
&\mathcal{L}_i w = \dfrac{\bar{\eta} \Omega}{4} \partial_p \left\lbrace i \spl e^{i \bar{\eta} x} + h.c., w \right\rbrace
\end{align}
As the ions are put into motion from the recoil kicks upon photon absorption, they will thus see a Doppler shift in the laser frequency.
This can be seen even more clearly by introducing a transformed Wigner-representation
\begin{equation}\label{eq:trafo-semi-polaron}
\tilde{w}(x,p) = e^{-i \bar{\eta} x \sz/2} w(x,p) e^{i \bar{\eta} x \sz/2}.
\end{equation}
Note that the Wigner function for the motional degree of freedom is still obtained as $ W(x,p) = \tr{\tilde{w}}$.
While transformation\,\eqref{eq:trafo-semi-polaron} is unitary with respect to the internal degree of freedom, it is a non-unitary transformation for the composite system, similar to the transformation introduced in\,\citep{LoerchQianClerkEtAl2014}. The Wigner function in the new representation obeys the equation of motion
\begin{equation}\label{eq:eomtildew}
\dot{\tilde{w}} = ( \mathcal{L}_m + \tilde{\mathcal{L}}_a + \tilde{\mathcal{L}}_i) \tilde{w}
\end{equation}
and the new operators are
\begin{align}
&\tilde{\mathcal{L}}_a \tilde{w} = -i \left[ -\dfrac{\Delta_{p}}{2} \sz + \dfrac{\Omega(t)}{2} \sx, \tilde{w} \right] + \dfrac{\Gamma}{2} \mathcal{D}\left[ \sm \right] \tilde{w} \label{eq:dynion} \\
&\tilde{\mathcal{L}}_i \tilde{w} = \dfrac{\bar{\eta} \Omega}{4} \partial_p \big( \sy \tilde{w} +\tilde{w} \sy \big) \label{eq:dyninteract}
\end{align}
where 
\begin{equation}\label{eq:DopplerDetuning}
    \Delta_{p} = \Delta - \bar{\eta} \wmode p
\end{equation}
is the Doppler-shifted detuning that the ion sees for a given momentum $p$.

During spectroscopy dynamics the spectroscopy ion is driven by short pulses where $\Omega(t)=\Omega$ for a time $\tpulse \ll T_m$. With each interaction a small photon recoil is imparted on the collective mode of motion. In this regime the internal dynamics will follow a comparatively fast dynamics (on the scale $\Omega,\Gamma$) generated by \eqref{eq:dynion} whose impact on the motional degree of freedom is at most on the order of $\bar\eta\Omega$. We therefore can adiabatically eliminate the dynamics of the internal degree of freedom and solve Eq.~\eqref{eq:eomtildew} for the Wigner function $W(x,p,t)$ in Eq.~\eqref{eq:Wigner} in first order of the Lamb-Dicke parameter $\eta$, as detailed in Appendix\,\ref{sec:app-OBE} (see also \citep{LoerchThesis2015}). This results in a relation
\begin{align}\label{eq:recursion}
W(x,p,t+\tpulse)=W(x,p,t)+\Delta W(x,p,\tpulse)
\end{align}
linking the Wigner function at the beginning and the end of each pulse. The increment $\Delta W(x,p,\tpulse)$ will be small (of order $\eta$). Each pulse is then followed by a trivial  oscillation of order $T_m$ generating essentially a $2\pi$ rotation of the Wigner function in phase space. Overall, relation \eqref{eq:recursion} can thus be interpreted as a recursion relation propagating the Wigner function from pulse to pulse. In view of the smallness of the increment $\Delta W$ and the large number of applied pulses (70 pulses were used in\,\cite{WanGebertWuebbenaEtAl2014}) it is justified to turn the difference equation \eqref{eq:recursion} into a differential equation for $W(x,p,\ttime)$ with a dimensionless spectroscopy time $\ttime=t/T_m$ counting the number of spectroscopy pulses (and subsequent $2\pi$ oscillations). The corresponding differential equation is of Fokker-Planck-type
\begin{equation}\label{eq:FokkerPlanck}
\dfrac{\partial}{\partial \ttime} W (x,p, \ttime) = \left[  \dfrac{\partial}{\partial p } \oneddrift(\Delta_{p})   + \dfrac{\partial^2}{\partial p^2} \dfrac{D(\Delta_{p})}{2} \right] W (x,p, \ttime).
\end{equation}
This result applies in zeroth order of $\tpulse\nu$, that is, neglecting the effect of oscillations during each pulse. The complete result is derived and discussed in Appendix\,\ref{sec:app-OBE}. In this limit the problem is reduced to the one-dimensional propagation in $p$-direction accounting for the photon recoil accumulated from all spectroscopy pulses up to time $\ttime$.

The Fokker-Planck equation is characterized by drift and diffusion coefficients, $\oneddrift(\Delta_p)$ and $D(\Delta_p)$ respectively. They follow from the solution of the optical Bloch equation for the internal dynamics  discussed in Appendix \ref{sec:app-OBE}. Their explicit form is  rather involved and will not be stated here. For the following discussion it is important to note that both coefficients depend on the Doppler-shifted detuning $\Delta_p$ \,(cf. Eq.~\eqref{eq:DopplerDetuning}) in the form of a Lorentzian resonance whose width is set by the probed transition.

\subsection{Frequency measurement}\label{subsec:Measurement}
\begin{figure}[tb]
\centering
\includegraphics[width=\linewidth]{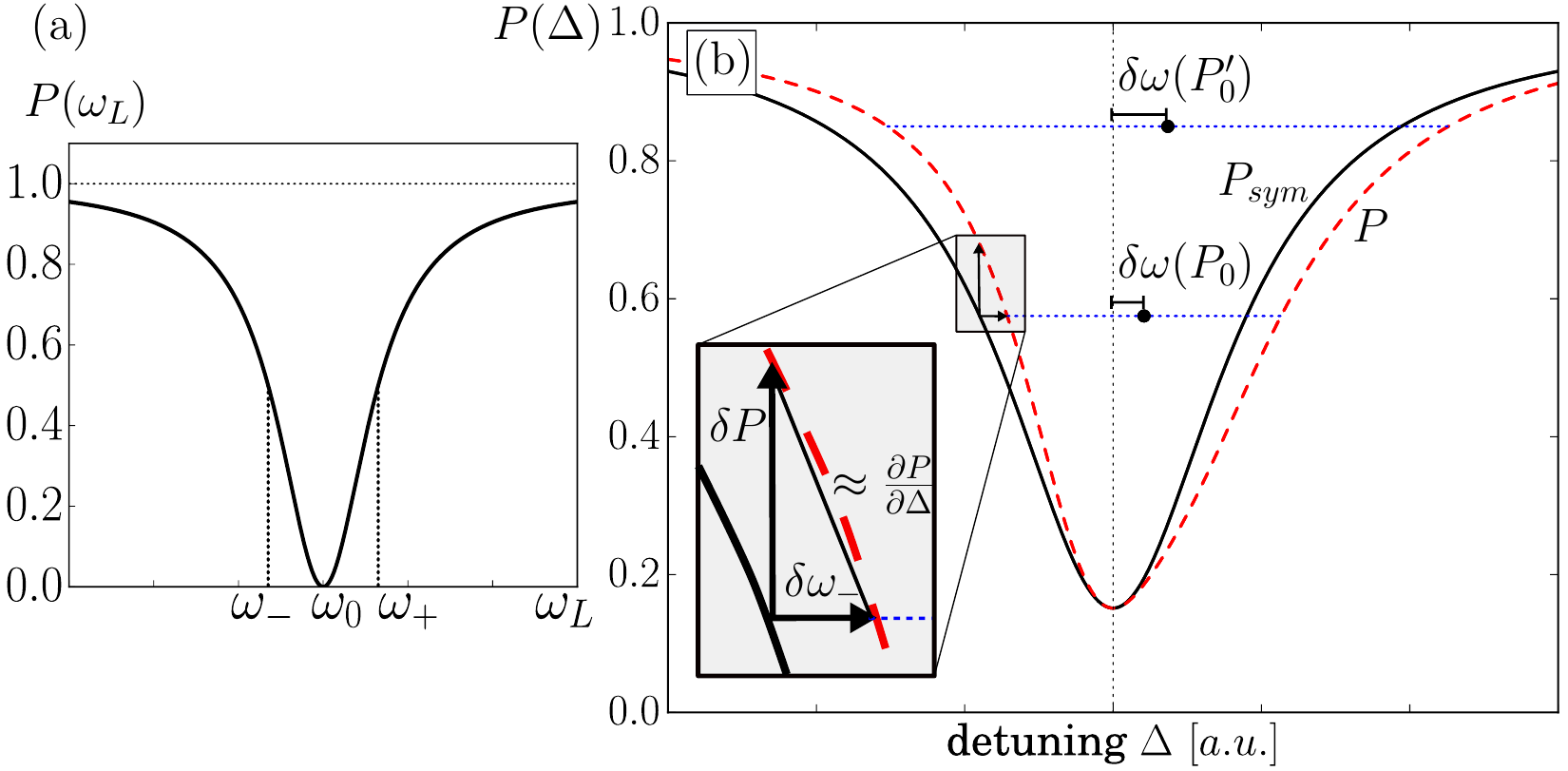}
\caption{(a) Schematic representation of two-point sampling. (b) Systematic Doppler-induced shift. The black, solid line depicts the symmetric resonance profile $P_{sym}$ obtained for $g=0$ and the red, dashed line depicts the true overlap $P$ which is shifted by $\delta P$ and therefore asymmetric around $\Delta=0$. The frequency shift $\delta \omega(P)$ is illustrated for two working points. The inset shows the linearisation of $P$ used to determine the shift on one side of the resonance.}
\label{fig:shift}
\end{figure}

So far we have only described the internal dynamics of the spectroscopy ion and the resulting evolution of the common motion.
To use this for spectroscopy of the transition frequency, a resonance profile must be measured and the corresponding central frequency has to be determined.
We assume that the motional mode was initially prepared in the pure state $\rho_0=\ket{\psi_0}\bra{\psi_0}$. The probability to remain in the initial state at time $\ttime$ is $P = \tr{\rho_0 \rho_{\ttime}}$. In the phase-space description of quantum optics adopted above, this can be calculated as an overlap integral\,\cite{Ferraro2005}
\begin{equation}
\label{eq:overlap} P = \pi \int_{\mathbb{R}^2} \mathrm{d}x\,\mathrm{d}p\, W_{[\rho_0]}(x, p)\, W_{[\rho_{\ttime}]}(x, p)
\end{equation}
between the Wigner functions for the two states. Here the Wigner function $W_{[\rho_{\ttime}]}(x, p)$ is the solution of the Fokker-Planck equation\,\eqref{eq:FokkerPlanck} with initial condition $ W_{[\rho_0]} $ after a time evolution $\ttime$.
This overlap probability depends on all parameters of the laser-ion interaction such as the spectroscopy time $\ttime$ and -- via the drift and diffusion coefficients -- on the detuning $\Delta$. For fixed $\ttime$ a qualitative illustration of the detuning dependence is shown in Fig.\,\ref{fig:shift}a, which features a characteristic dip at $\omega_L=\omega_0$. This is due to the fact that on resonance more photons are absorbed and the corresponding recoil leads to a large displacement of the initial state whereas only few photons will be absorbed with far-off resonant light so that the motion remains mostly unchanged and the overlap is near one.
For a symmetrical profile, the central frequency can be determined via two-point sampling of the laser frequency $\omega_L$ at the flanks of the resonance: The laser frequency is scanned on both flanks in order to determine the frequencies $\omega_\pm$ where the excitation probabilities match a suitably chosen working point $P_0=P(\omega_+) = P(\omega_-)$. The resonance frequency is then inferred as $\omega_0 = (\omega_- + \omega_+)/2$.

In the following sections we will use the formalism developed here to investigate two aspects of photon recoil spectroscopy: In Sec.~\ref{sec:Shift} we analyze systematic shifts in the determination of resonance frequencies resulting from an asymmetry in the excitation probability due to the Doppler effect. In Sec.~\ref{sec:GainInSpec} we elaborate on the statistical uncertainty with which the central frequency can be determined. This depends on the unavoidable quantum projection noise and the slope of the signal. Therefore, we will investigate how much the sensitivity of such measurements can be increased by nonclassical states despite the additional diffusion.

\section{Systematic Shift due to the Doppler Effect}\label{sec:Shift}

The drift and diffusion coefficients $\oneddrift(\Delta_p)$ and $D(\Delta_p)$ physically result, respectively, from the dipole force and the quantum fluctuations in the dipole force on the motion of the ions during each spectroscopy pulse. Accordingly, the dependence of $\oneddrift(\Delta)$ and $D(\Delta)$ on the bare detuning is of Lorentzian shape, that is even in $\Delta$. However the implicit dependence of both coefficients on the momentum coordinate $p$ via the Doppler-shifted detuning $\Delta_p$, cf. Eq.~\eqref{eq:dynion}, induces a small asymmetry to the resonance profile in the excitation probability, as we will explain now.
Considering only first order terms in the Lamb-Dicke parameter, we expand the drift coefficient as
\begin{equation}
\oneddrift(\Delta_p) = \oneddrift(\Delta)  - g(\Delta) p
\end{equation}
where we introduced
\begin{align}
g(\Delta)&=-\dfrac{\partial\, \oneddrift(\Delta)}{\partial \Delta}\dfrac{\partial\, \Delta_p}{\partial p}
=\bar{\eta} \wmode \dfrac{\partial\, \oneddrift(\Delta)}{\partial \Delta}.
\end{align}
This amounts to a constant, $p$-independent drift coefficient $\oneddrift$ and on top of that Doppler-induced damping or anti-damping depending on the sign of $g(\Delta)$. As $\alpha(\Delta)$ is an even function in the detuning, $g(\Delta)$ will be odd, so the sign changes from red to blue detuned laser light, i.\,e. $\Delta < 0$ and $\Delta > 0$ respectively. This will have two effects: Firstly, the displacement in momentum accumulated during a spectroscopy time $\ttime$ will be reduced or enhanced, depending on the sign of $g$. Secondly, the oscillator's phase space distribution will get squeezed (anti-squeezed) in momentum space. Both effects will contribute to the overlap integral in Eq.~\eqref{eq:overlap} generating an asymmetry in the measured resonance profile.
For small $g$, this effect can be evaluated by a series expansion of the propagator for the Fokker-Planck equation (see Appendix\,\ref{sec:app-FPE} for details).
Up to first order in $g$, the overlap probability is $P(\Delta) = P_{sym}(\Delta) + \delta P(\Delta) $, where $P_{sym}(\Delta)$ denotes the symmetric probability obtained for $g=0$ i.\,e. neglecting the Doppler effect, and
\begin{equation}\label{eq:deltaP}
\delta P(\Delta) = \dfrac{g(\Delta) \ttime}{2} c \, P_{sym}(\Delta)
\end{equation}
is then an odd function in $\Delta$, cf. Fig.\,\ref{fig:shift}b.
Here $c$ is a state dependent constant of order unity, as introduced in appendix \ref{sec:app-FPE}.

Therefore, evaluating the transition frequency via two-point sampling results in a systematic shift due to the Doppler effect, as indicated in Fig.~\ref{fig:shift}b for two different choices of the working point ($P_0,\,P'_0)$. The frequency shifts $\delta \omega_{\pm}$ for blue and red detuning can be quantitatively determined by the linear approximation
\begin{equation}
\dfrac{\partial P}{\partial \Delta} \simeq \pm \dfrac{\delta P}{\delta \omega_{\pm}}
\end{equation}
shown in the inset of Fig.\,\ref{fig:shift}b. With this it follows that $ \delta \omega_+ = \delta \omega_- $ so that the resonance frequency, measured via two-point sampling at a working point $P_0$, is shifted by
\begin{equation}\label{eq:DopplerShift}
 \delta \omega (P_0)  = - \dfrac{\delta P}{\partial P/\partial\Delta}\bigg|_{P_0}  = -\dfrac{g P_{sym}}{2} c \, \left( \dfrac{\partial P/\partial\Delta}{\ttime} \right)^{-1}\bigg|_{P_0}.
\end{equation}
For Gaussian states an analytic approximation on the size of $\delta \omega$ can be given and was used to calculate the Doppler-induced shift observed in\,\cite{WanGebertWuebbenaEtAl2014}.
There, the time evolution is completely determined by equations of motion for quadrature expectation values and the covariance matrix, as given explicitly in Appendix\,\ref{sec:app-FPE}.
Neglecting diffusion ($\oneddiff=0$) leads to the result:
\begin{equation}
 \delta \omega (\pin)  = \dfrac{\eta \nu}{2 \sqrt{\mathrm{ln}(1/\pin)}}
\end{equation}
where the connection $\oneddrift \ttime = \sqrt{2~ \mathrm{ln}(1/\pin)}$ between the drift coefficient, spectroscopy time and working point $ \pin $ was used.
Remarkably, this expression depends only on a few key quantities despite the rather involved actual dynamics.
Note that although the correction $g$ of the drift rate can itself be small, the resulting asymmetry may still cause a significant shift of the determined resonance frequency.
Especially at low excitation, i.e. $\pin$ close to $1$, the Doppler-induced shift is larger due to the low gradient of the resonance curve than for stronger excitation of the motion.
Finally, we would like to point out that the shift described here is not the regular Doppler shift of a moving ion.
The effect is rather an additional systematic of recoil spectroscopy by taking into account the Doppler effect contained in the dynamics.
The regular Doppler shift gives a uniform offset to the measured transition frequency which results in a broadening of the resonance for a distribution of velocities. This is in contrast to the distortion of the line profile found here.

\section{Quantum enhanced sensitivity under drift and diffusion}\label{sec:GainInSpec}

In the previous sections we introduced the principle of photon recoil spectroscopy and gave a theoretical description of the dynamics.
In the end, the examined transition frequency was determined from a two-point sampling on the flanks of the resonance curve.
In this section we will investigate how the sensitivity of this measurement can be increased by nonclassical states of motion.
We consider a process of drift and diffusion given by the Fokker-Planck equation
\begin{equation}\label{eq:Fokker-Planck-1D-nog}
\dfrac{\partial}{\partial \ttime} W_{[\rho]}(x, p, \ttime) = \left[ \oneddrift  \dfrac{\partial}{\partial p}+ \dfrac{\oneddiff}{2} \, \dfrac{\partial^2}{\partial p^2} \right] W_{[\rho]}(x, p, \ttime)
\end{equation}
with constant drift and diffusion coefficients $\oneddrift$ and $\oneddiff$ independent of the momentum variable $p$. This corresponds to Eq.\,\eqref{eq:FokkerPlanck} when neglecting the small Doppler-induced effects. We will discuss the impact of nonclassical enhancements on this systematic shift further below.

To understand the general effect of diffusion on the measurement, we focus at first on the case of a constant $\oneddiff$, independent of the detuning $\Delta$. This simplification will help to focus on our main interest here, that is the impact of diffusion on nonclassical enhancements. The spectroscopic signal is then only due to the drift $\alpha(\Delta)$. Due to the smallness of the influence of diffusion as compared to the drift during the spectroscopy time $\ttime$ the spectroscopic signal contributed by the actual $\Delta$-dependence of the diffusion is in any case relatively small, and will be analyzed later on.

\begin{figure}[tb]
\centering
\includegraphics[width=0.75\linewidth]{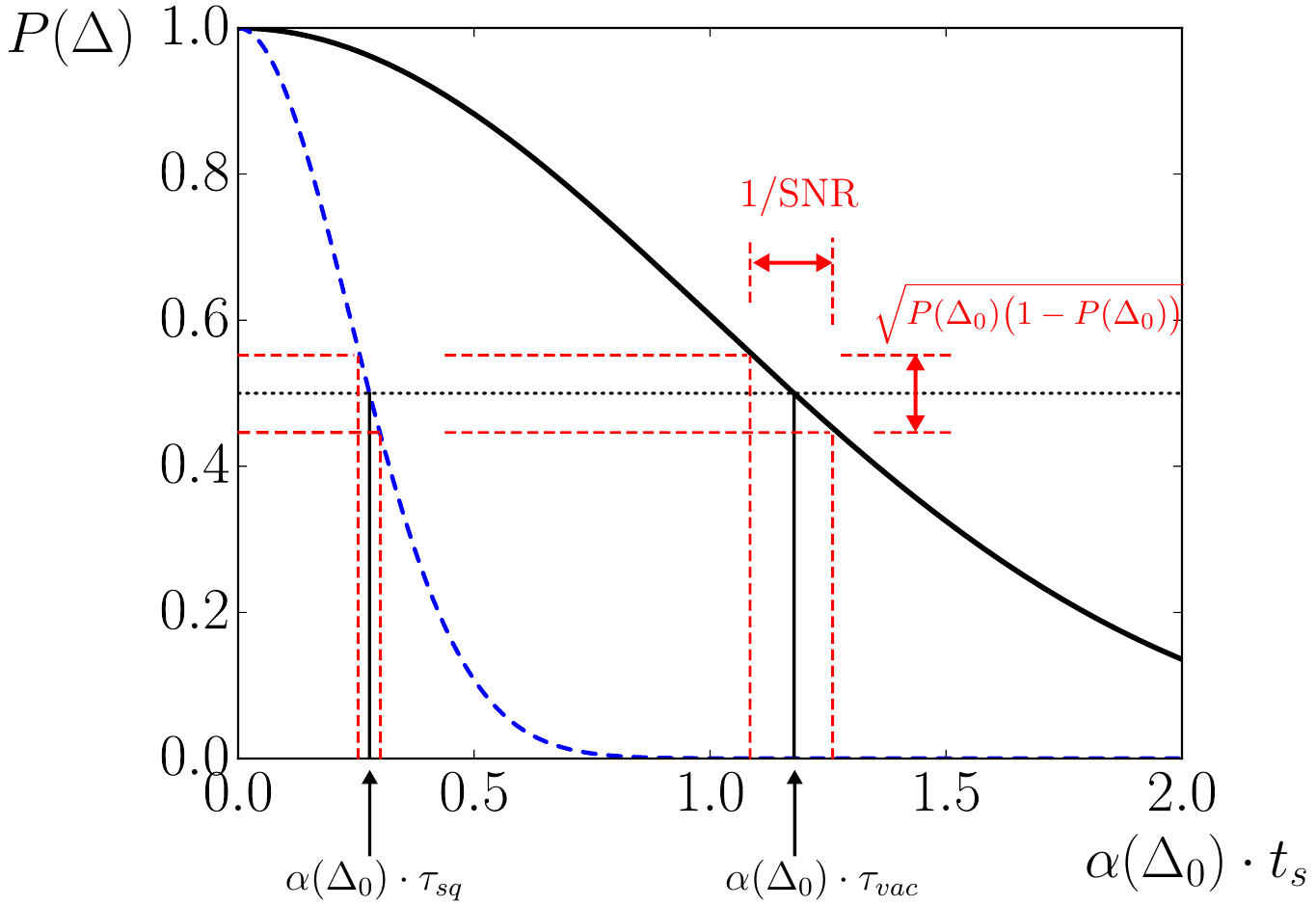}
\caption{Wigner function overlap $P$ for a momentum squeezed state (blue, dashed) and the ground state (black, solid) against displacement $\oneddrift(\para) \cdot \ttime$.
The signal-to-noise ratio depends on the quantum projection noise $\sqrt{P(1-P)}$ and the slope of $P$, which changes with the initial state $\rho_0$.
The interaction times $\tspec_{sq}$ and $\tspec_{vac}$ are defined to result in $\pin = 1/2 $ for both states.}
\label{fig:SNR}
\end{figure}

Intuitively, the quality of a measurement with excitation probability $P(\Delta)$ can be characterized by its signal-to-noise ratio achievable in $N$ measurements
\begin{equation}
    \mathrm{SNR} = \dfrac{\vert \partial P/\partial \Delta \vert_{\para}}{\sqrt{\pin (1-\pin)/N}}.
\end{equation}
As shown in Fig.\,\ref{fig:SNR}, this expresses the measurement uncertainty in $\Delta$ around the value $ \para $ corresponding to the chosen working point $\pin=P(\para)$ by the slope of $P(\Delta)$ at $\para$ and the projection noise $\sqrt{\pin(1-\pin)/N}$ of $N$ binary projective measurements. Usually the working point is chosen such that $P_0=1/2$ in order to maintain a large error signal. In Appendix \ref{sec:app-unitary} we show, by means of the Fisher information, that this working point becomes favourable if (infinitesimally small) imperfections in the binary measurement are taken into account. The fact that the projection noise becomes maximal at this point becomes irrelevant for large $N$.

Under this constraint, the $\mathrm{SNR}$ is essentially determined by the slope
\begin{equation}
\frac{\partial P}{\partial \Delta}\bigg|_{\para}=\frac{\partial P}{\partial \alpha}\bigg|_{\para}\frac{\partial\alpha}{\partial\Delta}\bigg|_{\para}.
\end{equation}
The dependence of the drift coefficient $\alpha$ on the detuning is set by the parameters of the spectroscopy pulses, and is independent of the choice of the quantum state of motion $\rho_0$. The steepest slope $\partial\alpha/\partial\Delta$ will be achieved for a detuning $\Delta_0$ on the scale of the width of the probed transition. In the following we will assume this factor to be fixed and optimized with respect to the relevant parameters including in particular the detuning $\para$. For a given detuning the spectroscopy time $\ttime=\tspec$ necessary to reach the working point $P_0$ then follows from
\begin{equation}
\label{eq:deftspec} P(\oneddrift(\para),\tspec)= \pin = 1/2.
\end{equation}
It is worth noting here that the required spectroscopy time $\tspec$ will depend on the initial state $\rho_0$ of motion, as illustrated by the example shown in Fig.\,\ref{fig:SNR}.

In applications of PRS, the probed transition will be typically broad and not closed. This makes the spectroscopy time $\ttime$ a precious resource as the number of absorbed photons increases linearly and with each scattering event there is a finite chance of loosing the electronic excitation to dark states. What is ultimately most relevant is, therefore, the SNR achievable per spectroscopy time, $\mathrm{SNR}/\tspec$.


Taking account of all of these aspects, we take as the measure of sensitivity in PRS the SNR per (dimensionless) spectroscopy time scaled by the constant coefficients introduced above, and define the \textit{recoil sensitivity} $S$ around $\para$ by
\begin{align}\label{eq:sensitivity}
S &= \dfrac{\mathrm{SNR}}{\tspec} \dfrac{\sqrt{\pin (1-\pin )/N}}{\vert \partial \oneddrift / \partial \Delta \vert_{\para}} = \dfrac{1}{\tspec} \dfrac{\partial P}{\partial \oneddrift}\bigg|_{\para}
\end{align}
where the interaction time $\tspec$ is the solution of Eq.~\eqref{eq:deftspec}. In the next sections we will study the dependence of the recoil sensitivity $S$ on the initial state $\rho_0$. We will discuss and compare various options for optimizing $S$ by using tailored quantum states, such as squeezed states, Fock states, or cat states. We will also discuss upper bounds on this quantity following from the Fisher information, and compare the results to this bound.

 \begin{figure*}[t]
\centering
\includegraphics[width=\linewidth]{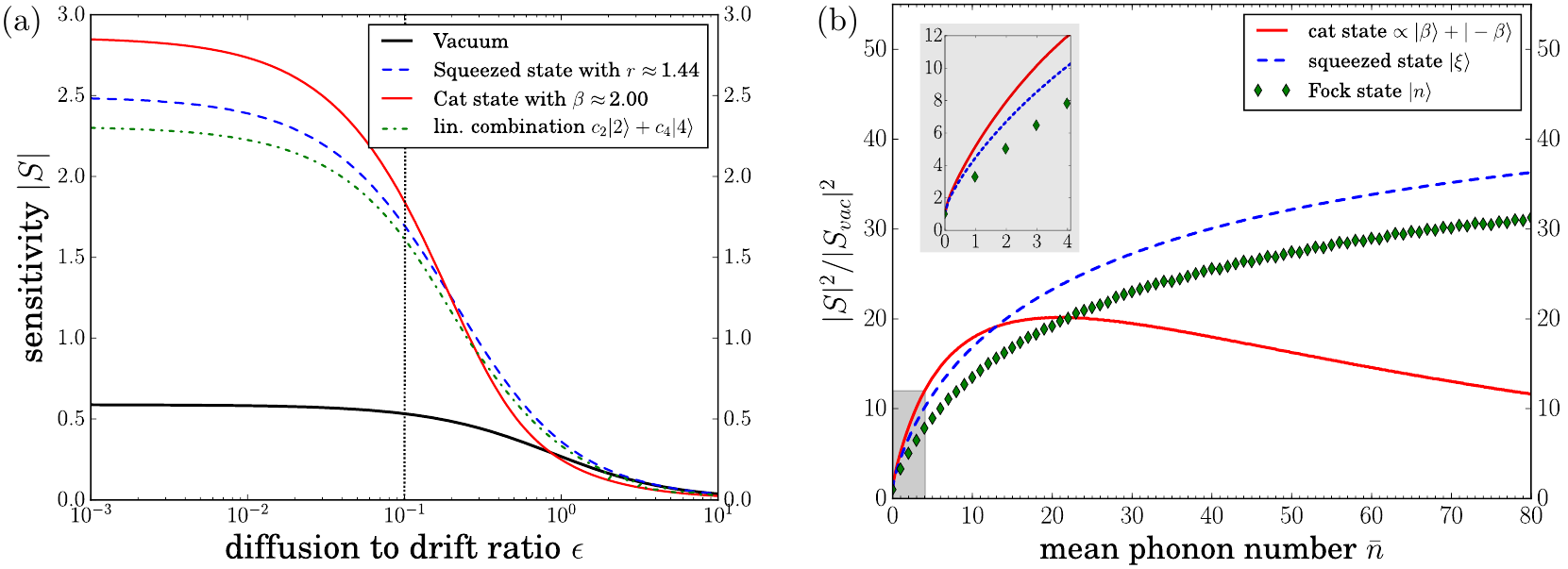}
\caption{(a)\,Recoil sensitivity under diffusive noise characterized by diffusion-to-drift ratio $\epsilon=D/\alpha$. The comparison shows the ground state $\ket{0}$\,(thick, black) along with a momentum squeezed state $\ket{r,\pi/2}$\,(blue dashed) and cat state $\propto \ket{\beta} + \ket{-\beta}$\,(red, full line), both constrained to $ \bar{n}=4 $. Additionally, optimized linear combinations of the first two even Fock states are considered (green, dash-dotted).
(b)\,Energy scaling of the sensitivity improvement over the ground state for the cat state, squeezed state and Fock states for fixed $\epsilon=D/\alpha=0.1$ as indicated in (a). The inset shows the approximate linear scaling at low $\bar{n}$.}
\label{fig:DD-IndependentDiffVgl}
\end{figure*}

\subsection{Sensitivity Enhancement with Constant Diffusive Noise}
The appeals of nonclassical probes can readily be seen for a selection of states that have shown large improvements under ideal displacements.
There, squeezed states $ \ket{r,\phi}=\exp\left[r/2 \left( a^2 e^{-2i\phi} - (a^{\dagger})^2 e^{2i\phi} \right) \right]\ket{0}$ with squeezing amplitude $r$ and phase $\phi$ were found to give an improvement by a factor of $ e^{2r} $ and Schr\"{o}dinger-cat states $ \ket{\mathrm{cat}} \propto \ket{\beta} + \ket{-\beta} $ with separation $2\beta$ between the coherent states $|\pm\beta\rangle$ improve by $ \approx \beta^2 $ (see Appendix\,\ref{sec:app-unitary}).
For completeness, the analytic solutions of the Fokker-Planck equation\,\eqref{eq:Fokker-Planck-1D-nog}, starting from arbitrary Gaussian states as well as from a Schr\"{o}dinger-cat state are given in Appendix\,\ref{sec:app-FPE}.
Based on these solutions, one finds the probability to remain in the initial state to be
\begin{equation}
\label{eq:Psq} P(\oneddrift, \oneddiff, \ttime, r) = \dfrac{1}{\sqrt{1+\oneddiff e^{2r} \ttime}} ~ \mathrm{exp}\left[ -\dfrac{1}{2}\dfrac{e^{2r} \oneddrift^2 \ttime^2}{1+\oneddiff e^{2r} \ttime} \right]
\end{equation}
for a momentum squeezed initial state $ \ket{\psi_0} =  \ket{r,\pi/2} $.
In the case of unitary displacement with $\oneddiff = 0$, this probability is decreasing as $ \mathrm{exp} \left[ - (e^r \oneddrift \ttime)^2/2 \right] $.
For $r=0$ this is the overlap between a coherent state with amplitude $ \oneddrift \ttime $ and the ground state, as would be expected.
The resulting sensitivity for the squeezed state according to Eq.\,\eqref{eq:sensitivity} is
\begin{align}
\label{eq:SforSq1} \vert S \vert &= \dfrac{1}{2} \dfrac{\oneddrift \tspec\, e^{2r}}{1+\epsilon\, \oneddrift\tspec\, e^{2r}}\bigg|_{\para} &
\epsilon &= \oneddiff /\oneddrift
\end{align}
where $\epsilon$ denotes the diffusion to drift ratio. A similar but more lengthy formula can be derived for the cat state and is given at the end of Appendix \ref{sec:app-FPE}.
Approximating Eq.\,\eqref{eq:SforSq1} for $ \epsilon \to 0 $ \footnote{In an unscaled phase space of physical units with coefficients $\tilde{\oneddrift}$ and $\tilde{\oneddiff}$ the limit $\epsilon \ll 1$ would be $\tilde{\oneddiff}/\tilde{\oneddrift} \ll p_{zpf}$ where $ p_{zpf}$ is the momentum zero point fluctuation of the mode.}, by first solving Eq.\,\eqref{eq:deftspec} in this limit, gives the diffusion-free result
\begin{equation}
\label{eq:SforP0approx} \vert S \vert \approx  e^r \sqrt{\mathrm{ln}(2)/2}
\end{equation}
which depends solely on the squeezing strength $r$ and shows Heisenberg scaling $\vert S\vert^2 \propto \overline{n}$, where $\overline{n}=\bra{\psi_0}a^\dagger a\ket{\psi_0}$ is the average occupation number of the initial state.
However analytic results beyond this limit, i.\,e. for arbitrary drift and diffusion, are not found as $\tspec$ itself depends non-analytically on $\oneddrift(\Delta)$ and $\oneddiff$ as well as the initial state $\rho_0$.

We note that for constant diffusion and drift coefficients their ratio $\epsilon$ defines equivalence classes of dynamics described by the Fokker-Planck equation \eqref{eq:Fokker-Planck-1D-nog} which will allow us to compare the gain achieved by different quantum states in dynamics characterized by a certain $\epsilon$.

As a main result of this study, the recoil sensitivity $\vert S \vert$ is calculated numerically as a function of the diffusion to drift ratio $\epsilon$ for different initial states $\ket{\psi_0}$. We focus here on states which have already been demonstrated in ion trap experiments\,\cite{MeekhofEtAl1996, MonroeEtAl1996, KienzlerEtAl2014, McCormicEtAl2018}.
The results are shown in Fig.\,\ref{fig:DD-IndependentDiffVgl}a, where we compare the following initial states: The ground state $\ket{\psi_0}= \ket{0}$ (black, thick line), a momentum squeezed state $\ket{\psi_0}= \ket{r,\pi/2}$ (blue dashed line), a cat state $\ket{\psi_0} \propto \ket{\beta} + \ket{-\beta}$ (red, thin line), and a numerically optimized linear combination of Fock states $\ket{\psi_0}= c_2 \ket{2}+c_4 \ket{4}$ (green, dash-dotted).
Note that the optimal coefficients $c_2$ and $c_4$ differ for different values of the diffusion to drift ratio $\epsilon$.  The corresponding line in Fig.\,\ref{fig:DD-IndependentDiffVgl}a does not arise from one single initial state but rather an optimal superposition for each ratio $\epsilon$.
See Appendix\,\ref{sec:app-numerics} for details on the numerical optimization.
In order to make a fair comparison between the states, we constrain their mean phonon number to $ \bar{n} = 4 $, which sets the squeezing parameter $r \approx 1.44$ and separation of coherent states in the cat state $\beta \approx 2 $.
However for the linear combination of Fock states the energy constraint is only an upper bound and some of the optimal states have $\bar{n}<4$.

Fig.\,\ref{fig:DD-IndependentDiffVgl}a shows that in this case the cat state gives the largest recoil sensitivity at small ratios $ \epsilon \ll 1 $, followed by the squeezed state and the linear combination, as it may be expected from results on unitary displacements (see Appendix\,\ref{sec:app-unitary}).
As the additional diffusion increases, each recoil sensitivity is reduced at individual rates with the largest effect on the cat state, eventually dropping below the other states.
In the limit of large diffusion to drift, i.e. $\epsilon \gtrsim 1$, all states result in almost equal sensitivities on the level of the ground state so that any possible quantum gain is lost there due to the dominating diffusion.

We also look at the $\bar{n}$ scaling of the sensitivity improvement over the ground state at fixed diffusion-to-drift ration $\epsilon=0.1$ in Fig.\,\ref{fig:DD-IndependentDiffVgl}b.
This shows representative curves of the scaling for the cat state, squeezed state and Fock states $\ket{n}$.
For non vanishing diffusion, so any $\epsilon>0$, the squeezed states will experience a saturation in an approximate regime where $\bar{n} \gg \epsilon^{-1}$. This can be inferred from Eq.\,\eqref{eq:SforSq1} for which $\oneddrift(\para) \ttime e^{2r}$ increases at most as $\sqrt{\bar{n}}$ but saturates for large $\bar{n}$ towards a constant value which increases with decreasing $\epsilon$.
A similar scaling feature is observed for the Fock states.
The cat states however are qualitatively different. They have the fastest scaling at small mean phonon numbers but then reach a maximum in the improved sensitivity before decreasing again. The fact that this feature also holds for arbitrary small diffusion can be inferred from the formula for $S$ in Appendix \ref{sec:app-FPE}, similarly to the squeezed state saturation.
All states have an approximate linear scaling for small $\bar{n}$ in common, as shown in the inset of Fig.\,\ref{fig:DD-IndependentDiffVgl}b.
The size for this regime of linear scaling increases with decreasing $\epsilon$ just like the onset of non-linear features. Unfortunately, however, no simple scaling law was found for these relations.


\begin{figure}[tb]
\centering
\includegraphics[width=\linewidth]{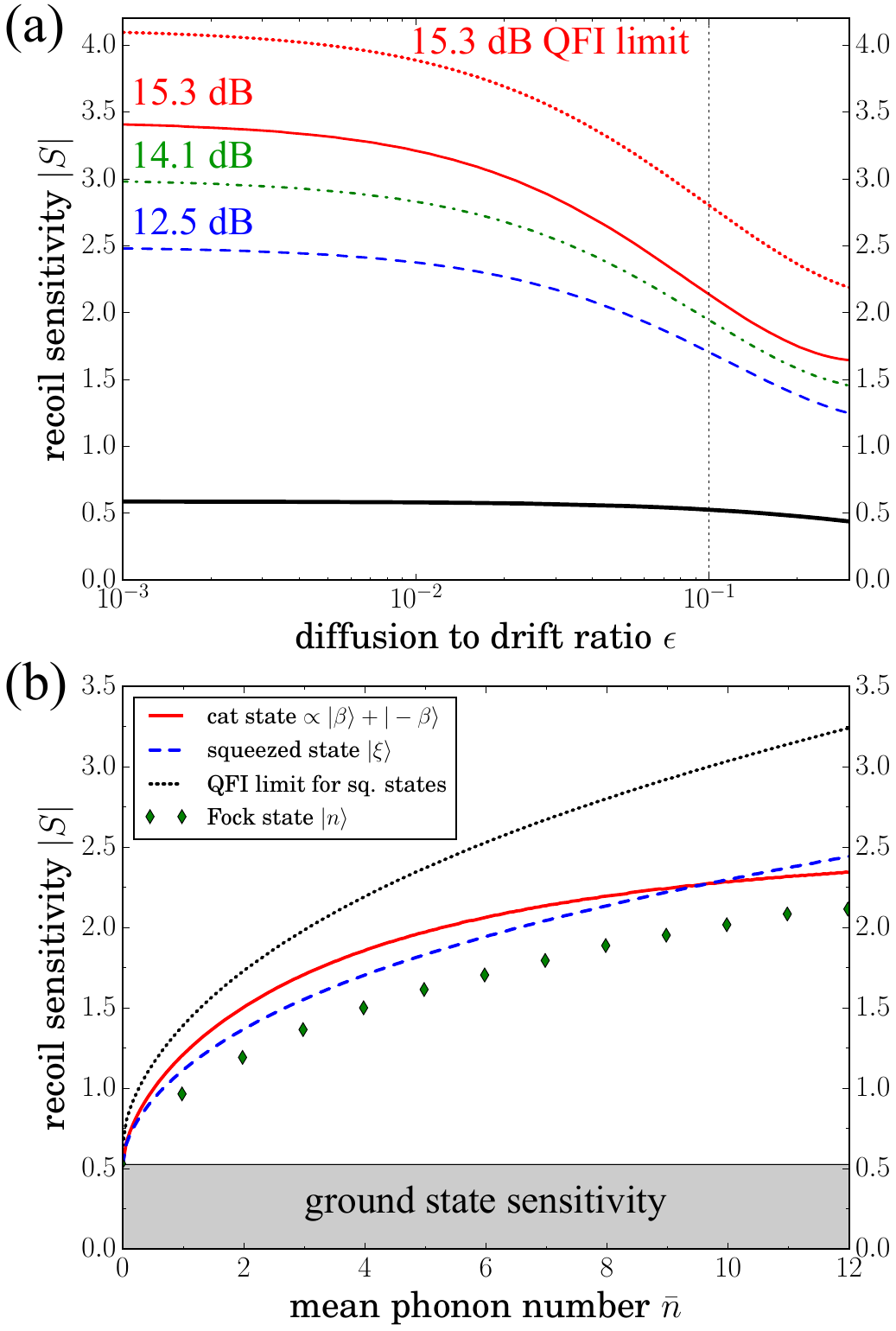}
\caption{(a) Recoil sensitivity for momentum squeezed states with $\bar{n} = 4, 6, 8 $\,(blue dashed, green dash-dotted, red solid) over the range of $\epsilon$. The ground state\,(black, thick line) is added as a reference. The red dotted line corresponds to the QFI sensitivity limit according to Eq.\,\eqref{eq:Sbound} for the state with $15.3\,\mathrm{dB}$ squeezing.
(b) Sensitivity as a function of mean occupation $ \bar{n} $ for a realistic diffusion to drift ratio $\epsilon=0.1$, as marked in\,(a). The black dotted line is again the QFI limit for squeezed states with the respective $\bar{n}$}
\label{fig:squeezed-states}
\end{figure}

\subsection{Sensitivity Enhancements in PRS and Fundamental Limitations}

The following segment makes now more realistic considerations about the drift and diffusion dynamics of PRS and discusses fundamental and technical limitations on the maximally achievable sensitivity.

First, the assumption that the diffusion $D$ does not depend on the detuning is now lifted. We assume instead that the ratio $ \oneddiff(\Delta)/\oneddrift(\Delta) \equiv \epsilon $ should be approximately constant over a small range of detunings around $ \para $. This assumption appears reasonable for PRS where almost all fluctuations are due to the laser-ion interaction and there is approximately no additional decoherence of the motion during the spectroscopy dynamics\,\cite{BrownnuttKumphRablEtAl2015}.
More formally, one can justify this by a connection between the time evolution for expectation values and two-time correlation functions due to the quantum regression theorem as applied in Appendix\,\ref{sec:app-OBE}.
Another constraint, specific to PRS, is that not all ratios $\epsilon$ of diffusion to drift should be considered.
The reason is that the drift is a first order process in the Lamb-Dicke parameter $\LD$ whereas the diffusion is of second order in $\LD$ and so far this was assumed to be a small parameter.
Therefore a restriction to the regime $ \epsilon \leq 0.3 $ is made based on $\LD \approx 0.1$, as considered in section \ref{sec:TheoreticalModel}.

Due to these adapted requirements the figure of merit, $S$, is extended to
\begin{equation}
\label{eq:DefS2}
    \vert S \vert = \dfrac{1}{\tspec} \left(\dfrac{\partial P}{\partial \oneddrift}\bigg|_{\para} + \epsilon \dfrac{\partial P}{\partial \oneddiff}\bigg|_{\para} \right)
\end{equation}
where for the second summand we used the assumption of constant $\oneddiff(\Delta)/\oneddrift(\Delta)$ around $\para$ to rewrite
\begin{equation}
    \dfrac{\partial \oneddiff(\Delta)}{\partial\Delta} = \epsilon \dfrac{\partial\oneddrift(\Delta)}{\partial\Delta}\,.
\end{equation}
The expression for the recoil sensitivity of e.g. a momentum squeezed state is then
\begin{equation}
 \vert S \vert = \dfrac{1}{2} \dfrac{e^{2r}}{1+\epsilon\, \oneddrift\tspec\, e^{2r}} \left( \oneddrift \tspec + \dfrac{\epsilon}{2} \dfrac{1 + e^{2r} \left( \epsilon\, \oneddrift\tspec - (\oneddrift \tspec)^2 \right)}{1+\epsilon\, \oneddrift\tspec e^{2r}}  \right)\bigg|_{\para}
\end{equation}
with the added term from the parameter dependence of the diffusion.

The results for the sensitivity in this case are shown exemplary for squeezed states in Fig.\,\ref{fig:squeezed-states}a.
As the correction scales with $\epsilon$, the result for $\bar{n}=4$ coincides very well with Fig.\,\ref{fig:DD-IndependentDiffVgl} on the side of small ratios $\epsilon$. Also for other states the sensitivity behaves very similar to the ones shown in Fig.\,\ref{fig:DD-IndependentDiffVgl}a with only small corrections for larger $\epsilon$ with the notable difference that in contrast to the previous discussion, the sensitivities will not decrease to the ground state level for larger $\epsilon$ as the diffusion now also carries information about the parameter.
Overall, the nonclassical states show an enhancement for the entire range of diffusion to drift considered as realistic for PRS.

We will now discuss fundamental upper bounds on the attainable enhancement in sensitivity for a given initial state, and make a comparison to what we have found in the previous section. In quantum metrology, a fundamental upper limit can be derived in terms of the Fisher information.
There, for determining an unknown parameter $\para$ from $N$ independent measurements, the Fisher information $F(\Delta_0)$ gives a lower limit to the variance of an unbiased estimator according to the Cram\'{e}r-Rao inequality\,\cite{BraunsteinCaves1994}
\begin{equation}
\mathrm{Var}(\Delta_{est})\big|_{\para} \geq \dfrac{1}{N\, F(\Delta_0)}.
\end{equation}
For a quantum mechanical system, the Fisher information can be maximized over all measurements on the state and this results in the so-called quantum Fisher Information (QFI) $\mathcal{F}$, which is ultimately determined only by the initial state $\ket{\psi_0}$ and the parameter dependency of the dynamics.
In the case of a binary measurement with probability $P(\Delta)$ the Fisher information is
\begin{equation}
    F(\para) = \dfrac{1}{P(\para) (1-P(\para))} \left(\dfrac{\partial P}{\partial \Delta}\bigg|_{\para} \right)^2\,.
\end{equation}
With this the recoil sensitivity
\begin{equation}
    S = \dfrac{1}{\tspec} \dfrac{\partial P}{\partial \Delta}\bigg|_{\para}\,\cdot \left(\dfrac{\partial \oneddrift}{\partial \Delta}\bigg|_{\para} \right)^{-1}
\end{equation}
can be expressed (using $\pin = 1/2$) as
\begin{equation}
    S = \dfrac{1}{2 \tspec} \left(\dfrac{\partial \oneddrift}{\partial \Delta}\bigg|_{\para} \right)^{-1} \sqrt{F(\para)}.
\end{equation}
So the maximum over any measurement is then
\begin{equation}\label{eq:Sbound}
    \vert S \vert = \dfrac{1}{2 \tspec} \left\vert\dfrac{\partial \oneddrift}{\partial \Delta}\bigg|_{\para} \right\vert^{-1} \sqrt{\mathcal{F}(\para)}
\end{equation}
with the QFI at $\para$.

For Gaussian states the QFI can be calculated analytically, see\,\cite{PinelJianTrepsEtAl2013}.
So a comparison of squeezed state sensitivities for different squeezing strength is displayed in Fig.\,\ref{fig:squeezed-states}a, along with the QFI limit.
The curves shown correspond to states with $\bar{n}=4$ ($\approx 12.5$\,dB squeezing), $\bar{n}=6$ ($\approx 14.1$\,dB) and $\bar{n}=8$ ($\approx 15.3$\,dB).
The red dotted line gives the maximally achievable sensitivity of the $15.3$\,dB squeezed state, based on Eq.\,\eqref{eq:Sbound}.
This shows that larger squeezing parameters lead to an increase in sensitivity but one also finds that the binary projective state measurement at $\pin =1/2$ is not ideal as seen by the discrepancy to the QFI limit.
Here, this is due to the choice of $\pin$, as for unitary dynamics the Fisher information of the projective measurement onto $\ket{\psi_0}$ actually reaches the QFI in the limit of vanishing displacement, i.\,e. $ \pin \rightarrow 1$.
However any loss of contrast will quickly reduce the sensitivity at this point as shown in Appendix\,\ref{sec:app-unitary}. In addition, the extreme points of the resonance profile do not produce a sufficient error signal, and are not suited to perform the two-point sampling.

The scaling of recoil sensitivity with respect to the invested resource in terms of mean phonon number $ \bar{n} $ is shown in part\,(b)\,of Fig.\,\ref{fig:squeezed-states}.
For this, we used $ \epsilon = 0.1 $ based on realistic parameters for a cycling dipole transition\,\citep{WanGebertWuebbenaEtAl2014}.
Comparing cat states, squeezed states and Fock states indicates that, starting from the ground state sensitivity, there is an immediate gain at small $ \bar{n} $ for all states and qualitatively similar results to Fig.\,\ref{fig:DD-IndependentDiffVgl}b are observed.
In addition, the calculated QFI bound displays an increasing gap to the respective squeezed states with larger $\bar{n}$ indicating that also the scaling of the sensitivity is changed under diffusion.

In addition to the fundamental bounds there are also technical limitations to the achievable sensitivity.
An example for this are preparation and measurement errors.
For squeezed states in particular, there can be a mismatch between the squeezing direction of the initial state and the projection state of the measurement.
In Appendix\,\ref{subsec:mismatch} it is shown that such errors reduce the sensitivity, especially for larger squeezing parameters $r$. Including technical imperfections gives rise to more realistic constraints on the range of useful squeezing.
Note that similar alignment requirements exist for the cat states.
Even though Fock states lead to the smallest recoil sensitivity in the comparison, they may actually be more useful in this regard. As they allow enhancements in any direction due to the rotational symmetry in phase space, whereas the squeezed and cat state are optimized for only one specific direction \cite{WolfShiHeipEtAl2018}.

Finally, we would like to add a comment on the effect of nonclassical states on the Doppler-induced shift.
When evaluating the shift $\delta \omega (\tilde{p})$ from Eq.\,\eqref{eq:DopplerShift} around $ \pin = 1/2$ one finds the following connection to the recoil sensitivity $S$
\begin{equation}
\vert \delta \omega (\pin) \vert = \dfrac{\vert g \vert \vert c \vert}{4} \dfrac{1}{\vert S \vert}
\end{equation}
where $c$ is a constant of order unity, cf. Eq.\eqref{eq:deltaP}.
Thus, the systematic shift scales inversely to the recoil sensitivity.
This scaling also results from the phase space description, where e.g.\,a squeezed state requires less displacement in momentum direction as the ground state to reach the same overlap $P$ and, since the Doppler detuning $\Delta_{\mathrm{dop}}$ changes linearly with $p$, the non classical states will therefore see a smaller change in the laser detuning due to their motion.
As the Doppler-induced shift can be one of the largest systematic shifts in photon recoil spectroscopy and its size will depend on the initial state of motion, good knowledge about the state preparation is necessary in order to compensate the observed asymmetries.
But because of the inverse scaling with $S$, using sensitivity enhancing states may reduce the Doppler-induced shift to a level where it is no longer a significant contribution in the error budged.

\subsection{Requirement for PRS with Single Photon Absorption}
We assess here the requirements to perform spectroscopy on a dipole transition as probed in\,\cite{WanGebertWuebbenaEtAl2014} with a single photon on average, meaning that one photon gives the displacement for $\pin \! = \! 1/2$ in the drift and diffusion model.
To compute this, consider the following relevant parameters for a two ion crystal of $\,^{25}\mathrm{Mg}^+$ and $\,^{40}\mathrm{Ca}^+$ to probe the $\,^{2}\mathrm{S}_{1/2} - \,^{2}\mathrm{P}_{1/2} $ transition of Calcium:
Lamb-Dicke parameter $ \LD = 0.108 $, normal mode frequency $ \wmode = 2 \pi \cdot 1.92\,\mathrm{MHz}$, Rabi frequency $ \Omega =  2 \pi \cdot 5.6\,\mathrm{MHz}$, linewidth $ \Gamma = 2 \pi \cdot 34\,\mathrm{MHz} $ and short spectroscopy pulses of duration $ \tpulse = 50\, \mathrm{ns} $ with a detuning $ \para = \Gamma/2 $ in order to reach $\pin = \! 1/2$ at the flank of the resonance profile.
Plugging these values into the optical Bloch equations\,\eqref{eq:OBE-driftP} and\,\eqref{eq:OBE-diffPP} from Appendix\,\ref{sec:app-OBE} results in the drift and diffusion coefficients $ \oneddrift \approx 2.03 \cdot 10^{-2} $, $ \oneddiff \approx 3.03 \cdot 10^{-3} $ and therefore a ratio $ \epsilon \approx 0.149 $.
Not only the drift and diffusion coefficients but also the average number of absorbed photons can be calculated from solutions to the optical Bloch equations.
A formal derivation of this statement was given in\,\cite{Cohen-Tannoudji_APINT_04} based on an argument we summarize in Appendix\,\ref{sec:app-average}.
Note that the analysis performed there does not rely on steady state dynamics which may not be reached during the short laser pulses.
From this one finds a spectroscopy interaction time $ \tspec \approx 7.068$, i.e. around seven pulses, necessary to absorb a single photon on average.
Finally, numerically solving $ P_{sq}(\oneddrift, \oneddiff, \tspec, r) = \pin = 1/2 $ for $r$ gives the squeezing strength needed to perform spectroscopy with a single photon on a squeezed state.
The result is $ r \approx 2.13$ ($\approx 18.5$\,dB squeezing), corresponding to a mean phonon number $ \bar{n} \approx 17.3 $. This can then in turn be used to calculate the necessary enhancement in $ \vert S \vert $ compared to the motional ground state, giving a required enhancement by a factor of $ \approx 8.5 $.
So in total, implementing PRS for a single absorption appears challenging under the parameters considered here.
However this is also largely due to the demanding constraint to reach the working point $\pin = 1/2$ with a single photon on average and there have been experiments with Schr\"{o}dinger cat states reporting the detection of single photon scattering without $ \pin = 1/2 $ \,\cite{HempelLanyonJurcevicEtAl2013}.
In order to reduce the requirements for squeezing, the operating point can be changed and thus also the probability $1-\pin$ to detect the signal of the photons. Alternatively, the Lamb-Dicke parameter can be increased.
However note that with reduced detection probability, i.e. $\pin>1/2$, on the one hand more data points have to be recorded, though this is not a fundamental limitation, and on the other hand the spectroscopy is more strongly restricted by imperfections of the measurement, see also Appendix \ref{sec:app-unitary}.
Exact knowledge of these effects would allow an optimization of $\pin$ with additional minimization of $r$, but goes beyond the scope of this article.

\section{Conclusion}\label{sec:conclusion}
In this article we have studied spectroscopy by the momentum transfer of photon absorptions with trapped ions.
We developed a theoretical model for the light-ion interaction during the laser pulses and showed that the effective spectroscopy dynamics of a common motional mode of a two-ion crystal can be described by a Fokker-Planck equation.
The characteristic variables are drift and diffusion of the phase space distribution, which result from the dipole force and fluctuations of this force.
Based on this result we were able to predict systematic shifts of the measured resonance frequency due to the Doppler effect and also calculate the strength of this shift.
Finally, we considered ways to increase recoil sensitivity through nonclassical states of motion in the case of diffusive noise and specifically for PRS.
The gain in sensitivity against the noise level as well as the scaling with the phonon number was analyzed and compared with a fundamental bound by the quantum Fisher information.

\paragraph*{Acknowledgments} We thank Yong Wan for helpful discussions.
We acknowledge funding from the DFG through CRC 1227 (DQ-mat), projects A06 and B05.

\bibliography{PRSTheory2}


\begin{appendix}
\begin{widetext}

\section{Adiabatic elimination and optical Bloch equations}\label{sec:app-OBE}

This section contains a short discussion of the adiabatic elimination and supplies details on solving the ions internal dynamics during spectroscopy pulses which gives drift and diffusion coefficients for the resulting Fokker-Planck equation.
A more detailed derivation of this is given in\,\cite{LoerchThesis2015}.

For weak interactions in a regime where $ \bar{\eta}\Omega \ll \wmode , \Gamma $ one can use a Dyson series to approximately solve the equations of motion\,\eqref{eq:dynmot}-\eqref{eq:dyninteract} for a single laser pulse by expanding up to second order in $\tilde{\mathcal{L}}_i$.
In an interaction picture with respect to $\tilde{\mathcal{L}}_a + \mathcal{L}_m$ this gives
\begin{align}
\tilde{w}(\tpulse) &= e^{(\tilde{\mathcal{L}}_a + \mathcal{L}_m) \tpulse} \tilde{w}(0) + \int_0^{\tpulse} \mathrm{d}t\, e^{(\tilde{\mathcal{L}}_a + \mathcal{L}_m) (\tpulse-t)} \tilde{\mathcal{L}}_i e^{(\tilde{\mathcal{L}}_a + \mathcal{L}_m) t} \tilde{w}(0)\nonumber\\
&+ \int_0^{\tpulse} \mathrm{d}t \int_0^{t} \mathrm{d}t' \, e^{(\tilde{\mathcal{L}}_a + \mathcal{L}_m) (\tpulse-t)} \tilde{\mathcal{L}}_i e^{(\tilde{\mathcal{L}}_a + \mathcal{L}_m) (t-t')}  \tilde{\mathcal{L}}_i e^{(\tilde{\mathcal{L}}_a + \mathcal{L}_m) t'} \tilde{w}(0)
\end{align}
for a pulse of duration $\tpulse$.
Calculating the Wigner function $W^{(n)}$ after pulse $n$ has to consider the free evolution of the motion, where the internal state is assumed to decay back to the ground state, and tracing out the internal state.
Performing this evaluation this leads to\,\cite{LoerchThesis2015}
\begin{equation}
W^{(n+1)}-W^{(n)} = -\dfrac{\bar{\LD}\Omega}{2}\int_0^{\tpulse} \mathrm{d}t\, \partial_{p_t} \left\langle \sigma_y (t) \right\rangle W^{(n)} + \dfrac{(\bar{\LD}\Omega)^2}{4}\int_0^{\tpulse} \mathrm{d}t \int_0^t \mathrm{d}t'\, \partial_{p_t} \partial_{p_{t'}} \mathrm{Re}\left[\left\langle \sigma_y (t)\sigma_y (t') \right\rangle\right] W^{(n)} \, . \label{eq:ToFPE}
\end{equation}
Here $ \partial_{p_t} = \partial_p \mathrm{cos}(\wmode t) - \partial_x \mathrm{sin}(\wmode t) $, and $\sigma_y (t)$ is the solution of the optical Bloch equations describing the internal dynamics.
As outlined in the main text, the recursion relation\,\eqref{eq:ToFPE} can be approximated by Gaussian dynamics which gives the general Fokker-Planck equation
\begin{equation}\label{eq:FokkerPlanck-general}
\dfrac{\partial}{\partial \ttime} W (\textbf{x}, \ttime) = \left[ \sum_i \dfrac{\partial}{\partial \textbf{x}_i } \textbf{\gendrift}_i (\textbf{x})   + \sum_{i,j} \dfrac{\partial^2}{\partial \textbf{x}_i \partial \textbf{x}_j} \dfrac{\gendiff_{ij}(\textbf{x})}{2} \right] W (\textbf{x}, \ttime)
\end{equation}
describing the time evolution of the Wigner function $W(\textbf{x}, \ttime)$ by a drift vector $ \textbf{\gendrift} (x,p) $ and symmetric diffusion matrix $ \gendiff_{ij}(x,p) $.
Here $ \textbf{x} = \left(x, p\right)^\intercal $ and $\ttime = t/T_m $ is a dimensionless interaction time, which can be interpreted as a continuous extension for the number of applied spectroscopy pulses of duration $\tpulse$.
The drift and diffusion coefficients of Eq.\,\eqref{eq:FokkerPlanck-general} are obtained by calculating expectation values and correlation functions of $\sy$.
This can be done for the dynamics of $\tilde{\mathcal{L}}_a$ from Eq.\,\eqref{eq:dynion} via the optical Bloch equations
\begin{equation}
\dfrac{\mathrm{d}}{\mathrm{d}t} \left\langle \vec{\sigma}(t) \right\rangle = M \left\langle \vec{\sigma}(t) \right\rangle + \Gamma \vec{m}
\end{equation}
with
\begin{equation}
M = \begin{pmatrix}
-\dfrac{\Gamma}{2} & -\Delta_{\mathrm{dop}} & 0\\
\Delta_{\mathrm{dop}} & -\dfrac{\Gamma}{2} & -\Omega\\
0 & \Omega & -\Gamma
\end{pmatrix}
\hspace{2cm}
\vec{m} = - \begin{pmatrix}
0\\
0\\
1
\end{pmatrix}
\end{equation}
The solution for an initial condition $ \left\langle \vec{\sigma}(0) \right\rangle = \vec{m} $ is then
\begin{equation}
\left\langle \vec{\sigma}(t) \right\rangle = \left[ e^{M t} \left(1 + \Gamma M^{-1} \right) - \Gamma M^{-1} \right]\vec{m}.
\end{equation}
Calculating two-time correlation functions follows from the quantum regression theorem as
\begin{equation}
\label{eq:regressionTheorem} \dfrac{\mathrm{d}}{\mathrm{d}t} \left\langle \vec{\sigma}(t+t')\, \sy(t) \right\rangle = M \left\langle \vec{\sigma}(t+t')\, \sy(t) \right\rangle + \left\langle \vec{\sigma}(t) \right\rangle \Gamma \vec{m}
\end{equation}
with the initial condition
\begin{equation}
\left\langle \vec{\sigma}(t)\, \sy(t) \right\rangle = \begin{pmatrix}
i \left\langle \sz(t) \right\rangle\\
1\\
-i \left\langle \sx(t) \right\rangle
\end{pmatrix}
\end{equation}
For this, the solution is then given by
\begin{equation}
\left\langle \vec{\sigma}(t+t')\, \sy(t) \right\rangle = e^{M t'} \left( \left\langle \vec{\sigma}(t)\, \sy(t) \right\rangle + \left\langle \sy(t) \right\rangle \Gamma M^{-1} \vec{m} \right) - \left\langle \sy(t) \right\rangle \Gamma M^{-1} \vec{m}.
\end{equation}
In total, this results in the following drift and diffusion coefficients for the Fokker-Planck equation:
\begin{align}
\gendrift_x &= \dfrac{\eta \Omega}{\sqrt{2}} \int_0^{\tpulse} \mathrm{d}t\, \mathrm{sin}(\wmode t) \left\langle \sy (t) \right\rangle \\
\label{eq:OBE-driftP} \gendrift_p &= - \dfrac{\eta \Omega}{\sqrt{2}} \int_0^{\tpulse} \mathrm{d}t\, \mathrm{cos}(\wmode t) \left\langle \sy (t) \right\rangle \\
\gendiff_{xx} &= \eta^2 \Omega^2 \int_0^{\tpulse} \mathrm{d}t \int_0^t \mathrm{d}t'\, \mathrm{sin}(\wmode t)\, \mathrm{sin}(\wmode t') \mathrm{Re}\left( \left\langle \sy (t) \sy (t') \right\rangle \right) - \gendrift_x^2 \\
\label{eq:OBE-diffPP} \gendiff_{pp} &= \eta^2 \Omega^2 \int_0^{\tpulse} \mathrm{d}t \int_0^t \mathrm{d}t'\, \mathrm{cos}(\wmode t)\, \mathrm{cos}(\wmode t') \mathrm{Re}\left( \left\langle \sy (t) \sy (t') \right\rangle \right) - \gendrift_p^2 \\
\gendiff_{xp} &= -\eta^2 \Omega^2 \int_0^{\tpulse} \mathrm{d}t \int_0^t \mathrm{d}t'\, \dfrac{1}{2} \left[ \mathrm{sin}(\wmode t) \mathrm{cos}(\wmode t') - \mathrm{cos}(\wmode t) \mathrm{sin}(\wmode t') \right] \mathrm{Re}\left( \left\langle \sy (t) \sy (t') \right\rangle \right) - \gendrift_x \gendrift_p
\end{align}

Note that for spectroscopy pulses short on the timescale of a mechanical oscillation, one can approximate $ \gendrift_x = \gendiff_{xx} = \gendiff_{xp} = 0 $ since they are reduced by at least a factor $ \wmode \tpulse \ll 1$ compared to their respective momentum counterparts. With this approximation, the Fokker-Planck equation\,\eqref{eq:FokkerPlanck-general} reduces to the one-dimensional case
\begin{equation}
\label{eq:Fokker-Planck-1D} \dfrac{\partial}{\partial \ttime} W_{[\rho]}(x, p, \ttime) = \left[ \dfrac{\partial}{\partial p} \oneddrift(\Delta_p) + \dfrac{\partial^2}{\partial p^2} \dfrac{\oneddiff(\Delta_p)}{2} \right] W_{[\rho]}(x, p, \ttime)
\end{equation}
stated in the main text with the shorthand notations $\gendrift_p \equiv \oneddrift $ and $\gendiff_{pp} \equiv \oneddiff$.

\section{Solving the drift and diffusion Fokker-Planck equation}\label{sec:app-FPE}

In this section we revise methods to solve Fokker-Planck equations of the form:
\begin{equation}
\dfrac{\partial}{\partial \ttime} W (\textbf{x}, \ttime) = \left[ \sum_i \alpha_i \dfrac{\partial}{\partial \textbf{x}_i } + \sum_{l,m} \beta_{lm} \dfrac{\partial}{\partial \textbf{x}_l} \textbf{x}_m + \sum_{l,m} \dfrac{1}{2} \gendiff_{lm} \dfrac{\partial^2}{\partial \textbf{x}_l \partial \textbf{x}_m} \right] W (\textbf{x}, \ttime) \label{eq:FPappendix}
\end{equation}
with $ \textbf{x} = \left(x, p\right)^\intercal $ and specify on analytic results for the class of Gaussian states as used in the main text.
Solutions to Eq.\,\eqref{eq:FPappendix} can generally be obtained by the methods of Fourier transform and Characteristic functions or using a propagator function\,\citep{Risken1989}.
Any Wigner function for a density matrix $ \rho $ can be expressed in terms of its characteristic function via Fourier transform:
\begin{equation}\label{eq:Wigner-char}
W_{\left[\rho \right]} (\mathbf{x}) = \int \dfrac{\mathrm{d}^{2} \mathbf{k}}{\left( 2 \pi \right)^{2}} \, \mathrm{exp} \left[i\, \textbf{k}^\intercal\!\! \cdot \textbf{x} \right] \, \chi_{\left[\rho \right]} (\mathbf{k})
\end{equation}
The characteristic function is given by
\begin{equation}
\chi_{\left[\rho \right]} (k_1,k_2) = \tr{\rho\, \hat{D}(k_1, k_2)}
\end{equation}
with the displacement operator $ \hat{D}(k_1, k_2) = \mathrm{exp}\bigg[\big((k_1+i k_2) a + (k_1-i k_2) a^{\dagger}\big)/\sqrt{2}\bigg] $.
Gaussian states are described by characteristic functions of the form\,\citep{Ferraro2005}
\begin{equation}\label{eq:gaussianChar}
\chi (\textbf{k}) = \mathrm{exp} \left[ -\dfrac{1}{2} \textbf{k}^\intercal \gamma (\ttime) \textbf{k} + i\, \textbf{k}^\intercal\!\! \cdot \textbf{M} (\ttime) \right]
\end{equation}
with time dependent displacement vector $ \mathbf{M} (\ttime) = \langle \hat{\mathbf{R}} \rangle $ and covariance matrix $ \gamma_{ij} (\ttime) = \langle  \hat{R}_i \hat{R}_j + \hat{R}_j \hat{R}_i \rangle - 2 M_i M_j $, where $\mathbf{R} = (\hat{x}, \hat{p})^\intercal$ describes the quadrature operators $\hat{x} = \big(a + a^{\dagger}\big)/\sqrt{2}$ and $\hat{p} = -i\big(a - a^{\dagger}\big)/\sqrt{2}$.
The dynamics of the Fokker-Planck equation then leads to differential equations
\begin{align}
\dot{M}_i (\ttime) &= \dx M_j (\ttime) + \dc \label{eq:FP-EOM1} \\
\dot{\gamma}_{ij} (\ttime) &= \beta_{il} \gamma_{lj} + \beta_{il} \gamma_{jl} + \gendiff_{ij}\label{eq:FP-EOM2}
\end{align}
on these first and second moments\,\citep{Risken1989}.\\
Based on the Doppler effect discussed in the main text, we assume $ \beta = \begin{pmatrix}
0 & 0\\
0 & -g
\end{pmatrix} $, simplifying the equations of motion to
\begin{align*}
\dot{M}_x (\ttime) &= \alpha_x \\
\dot{M}_p (\ttime) &= -g M_p (\ttime) +\alpha_p \\
\dot{\gamma}_{xx} (\ttime) &= \gendiff_{xx} \\
\dot{\gamma}_{xp} (\ttime) &= \dot{\gamma}_{px} (\ttime) = -g \gamma_{xp} (\ttime) + \gendiff_{xp} \\
\dot{\gamma}_{pp} (\ttime) &= -2 g \gamma_{pp} (\ttime) + \gendiff_{pp}
\end{align*}
For initial conditions $ \textit{\textbf{M}}\,(0) = \begin{pmatrix}
x_{init}\\
p_{init}
\end{pmatrix} $ and $ \gamma (0) = \begin{pmatrix}
\gamma^0_{xx} & \gamma^0_{xp}\\
\gamma^0_{xp} & \gamma^0_{pp}
\end{pmatrix} $
the solutions are
\begin{align*}
M_x (\ttime) &= x_{init} +\alpha_x \ttime \\
M_p (\ttime) &= p_{init}\, e^{-g \ttime} + \dfrac{1}{g} (1-e^{-g \ttime}) \alpha_p \\
\gamma_{xx} (\ttime) &= \gamma^0_{xx} + \gendiff_{xx} \ttime \\
\gamma_{xp} (\ttime) &= \gamma_{px} (\ttime) = \gamma^0_{xp} e^{-g \ttime} + \dfrac{1}{g} (1-e^{-g \ttime}) \gendiff_{xp} \\
\gamma_{pp} (\ttime) &= \gamma^0_{pp} e^{-2 g \ttime} + \dfrac{1}{2 g} (1-e^{-2 g \ttime}) \gendiff_{pp}
\end{align*}
The dynamics discussed in the main text is approximated by $\alpha_x = 0$, $\alpha_p = \oneddrift$, $\gendiff_{xx} = \gendiff_{xp} =\gendiff_{px} =0$ and $\gendiff_{pp} = \oneddiff $.
So for the momentum squeezed vacuum state with $ x_{init} = p_{init} = 0 $ and covariance $ \gamma_{xx}^0 = \dfrac{1}{2} e^{2r}$, $\gamma_{pp}^0 = \dfrac{1}{2} e^{-2r} $, and $\gamma_{xp}^0 = 0$ gives:
\begin{align}
M_x (\ttime) &= 0 \label{eq:sqEOMsol1} \\
M_p (\ttime) &= \dfrac{1}{g} (1-e^{-g\ttime}) \oneddrift \approx \oneddrift \ttime - \dfrac{g \oneddrift}{2} \ttime^2 \label{eq:sqEOMsol2} \\
\gamma_{xx} (\ttime) &= \dfrac{1}{2} e^{2r} \label{eq:sqEOMsol3} \\
 \gamma_{xp} (\ttime) &= 0 \label{eq:sqEOMsol4} \\
\gamma_{pp} (\ttime) &= \dfrac{1}{2g} (1-e^{-2g \ttime}) \oneddiff + \dfrac{1}{2} e^{-2r} e^{-2g \ttime} \label{eq:sqEOMsol5} \\
&\approx \oneddiff \ttime + \dfrac{1}{2} e^{-2r} - g \ttime \left( \oneddiff \ttime + e^{-2r} \right) \label{eq:sqEOMsol6}
\end{align}
Then the overlap probability reduces to a Gaussian integral, as
\begin{align}
P &= \tr{\rho_0 \rho_{\ttime}} = \int_{\mathbb{R}^2}\dfrac{\mathrm{d}^2\mathbf{k}}{2\pi} \chi(\mathbf{k},\ttime=0) \chi(-\mathbf{k},\ttime) \\
&= \int_{\mathbb{R}^2}\dfrac{\mathrm{d}^2\textbf{k}}{2\pi} \mathrm{exp} \left[ -\dfrac{1}{2} \textbf{k}^\intercal \left( \gamma(0)+\gamma(\ttime) \right) \textbf{k} + i \left( \textit{\textbf{M}}(0)-\textit{\textbf{M}}(\ttime)\right)^\intercal \cdot \textbf{k} \right]\\
&= \dfrac{1}{\sqrt{\mathrm{det}(\gamma(0)+\gamma(\ttime))}} \mathrm{exp}\left[-\dfrac{1}{2} \left( \textit{\textbf{M}}(0)-\textit{\textbf{M}}(\ttime)\right)^\intercal \left( \gamma(0)+\gamma(\ttime) \right)^{-1} \left( \textit{\textbf{M}}(0)-\textit{\textbf{M}}(\ttime) \right) \right]
\end{align}

In contrast to the previously outlined method, solutions can also be obtained by determining the propagator for quasi-probability distributions.
The time evolution of a Wigner function starting from $\ttime=0$ is given by:
\begin{equation}
W(x,p,\ttime) = \int_{-\infty}^{\infty}\mathrm{d}x' \int_{-\infty}^{\infty}\mathrm{d}p' \, \mathcal{P}(x,p,t \vert x',p',0)\, W(x',p',0)
\end{equation}
Where the propagator for the case of PRS is
\begin{equation}
\mathcal{P}(x,p,t \vert x',p',0) = \sqrt{\dfrac{1}{2\pi \, \oneddiff/2g (1-e^{-2g\ttime})}} \mathrm{exp}\left[ - \dfrac{(p-e^{-g\ttime}p' - \oneddrift/g(1-e^{-g\ttime}))^2}{\oneddiff/g(1-e^{-2g\ttime})} \right] \delta (x-x')
\end{equation}
in analogy to a shifted Ornstein-Uhlenbeck process\,\citep{Risken1989}.
For small Doppler effects, this propagator can be expanded in a series around $g=0$. To first order one finds:
\begin{equation}\label{eq:propagator}
\mathcal{P}(x,p,t \vert x',p',0) = \mathcal{P}\bigg|_{g=0} + \dfrac{g \ttime}{2} \mathcal{P}\bigg|_{g=0} + \dfrac{g \ttime}{2} \dfrac{1}{\oneddiff \ttime} \left(-p^2 +p'^2 + \oneddrift \ttime p + \oneddrift \ttime p' \right) \mathcal{P}\bigg|_{g=0} + \mathcal{O}(g^2)
\end{equation}
so that the overlap probability is approximated by $ P = P_{sym} + \dfrac{g \ttime}{2} c \cdot P_{sym} $ with a state dependent constant $c$ of order unity that arises from the second and third third term on the right hand side of Eq.\eqref{eq:propagator}.

In addition to purely Gaussian states we also consider superpositions of such states, namely in the form of Schr{\"o}dinger cat states.
For a cat state $ \ket{\psi}= \mathcal{N} \left( \ket{\beta} + \ket{-\beta} \right) $ with $ \beta \in \mathbb{R} $ and normalization $ \mathcal{N} =  (2+2\,e^{-2\vert \beta \vert^2})^{-1/2} $, the characteristic function is:
\begin{equation}
\chi (k_1, k_2, \ttime=0) =  2\,\mathcal{N}^2 e^{-(k_1^2 + k_2^2)/4} \left( \mathrm{cos}(\sqrt{2} k_2 \beta) + e^{-2 \beta^2} \mathrm{cosh}(\sqrt{2} k_1 \beta) \right)
\end{equation}
The time evolvion for $g=0$ is given by:
\begin{equation}
\chi (k_1, k_2, \ttime) = \chi (k_1, k_2, \ttime=0)\, e^{i k_2 \oneddrift \ttime - k_2^2 \ttime \oneddiff/2}
\end{equation}
which leads to the overlap probability
\begin{equation}
P_{\mathrm{cat}} = \dfrac{1}{2} \dfrac{\sqrt{1/(1+\oneddiff \ttime)}}{(1+e^{-2 \beta^2})^2} e^{-\dfrac{\oneddrift^2 \ttime^2/2}{1+\oneddiff \ttime}} \left(1+2 e^{-4\beta^2}+e^{-\dfrac{4\oneddiff \ttime \beta^2}{1+\oneddiff \ttime}} \mathrm{cos}\left(\dfrac{2\sqrt{2} \oneddrift \ttime \beta}{1+\oneddiff t}\right) + 4\, e^{-\dfrac{\beta^2 (2+3\oneddiff \ttime)}{1+\oneddiff \ttime}} \mathrm{cos}\left(\dfrac{\sqrt{2}\oneddrift \beta \ttime}{1+\oneddiff \ttime}\right)\! \right)\, .
\end{equation}
This results in the recoil sensitivity
\begin{align*}
    \vert S \vert = \dfrac{1}{2} \dfrac{\oneddrift \tspec}{1 + \epsilon \oneddrift \tspec} &+ \dfrac{4 \mathcal{N}^4}{\sqrt{1 + \epsilon \oneddrift\tspec}} e^{- \dfrac{\oneddrift^2 \tspec^2/2}{1+\epsilon \oneddrift\tspec}} \dfrac{\sqrt{2}\beta}{1 + \epsilon \oneddrift\tspec} \Bigg( e^{-\dfrac{4 \oneddiff \tspec \beta^2}{1+ \epsilon \oneddrift\tspec}} \sin\Bigg( \dfrac{2 \oneddrift \tspec \sqrt{2} \beta}{1+ \epsilon \oneddrift\tspec}\Bigg)\\
    &+ 2 e^{-\dfrac{(2+3\epsilon \oneddrift\tspec)\beta^2}{1+\epsilon \oneddrift\tspec}} \sin\Bigg( \dfrac{\oneddrift \tspec \sqrt{2} \beta}{1+\epsilon \oneddrift\tspec}\Bigg) \Bigg)\bigg|_{\para}
\end{align*}
for $\oneddiff \neq \oneddiff(\Delta)$ and with $\epsilon = \oneddiff/\oneddrift$.

\section{Quantum metrology with unitary displacements and measurement imperfections}\label{sec:app-unitary}

A special case of the drift and diffusion process described by the Fokker-Planck equation\,\eqref{eq:Fokker-Planck-1D-nog} of the main text are the unitary displacements, i.e. when $ \oneddiff = 0 $.
This section reviews results in this case and includes a discussion about the influence of measurement imperfections.
We consider the standard scenario of quantum metrology as considered for an ideal interferometer: A quantum system, initially prepared in a state $\rho_0 = \vert \psi_0 \rangle \langle \psi_0 \vert$, evolves under a displacement in momentum direction with the unitary evolution
\begin{equation}\label{eq:App_Displ}
U = e^{i \oneddrift \ttime \hat{x}}
\end{equation}
where $ \hat{x} =\big(a + a^{\dagger}\big)/\sqrt{2} $ is the generator of the dynamics based on the canonical commutation relation between position and momentum operators.
The extend $\theta = \oneddrift \ttime$ of this displacement is to be estimated through a suitable measurement on the evolved state.
Quantum metrology treats this problem within the framework of parameter estimation where the Fisher information provides a measure for the quality of an estimation strategy.
The Fisher information $F$ gives a lower bound on the variance of an estimator $\theta_{est}$ in the form of a Cram\'{e}r-Rao bound\,\cite{BraunsteinCaves1994}
\begin{equation}
\mathrm{Var}(\theta_{est}) \geq (N\, F)^{-1}
\end{equation}
which holds for any unbiased estimation strategy with $N$ independent trials.
In particular, by optimizing over all estimators and measurements (described by POVM elements $\mathcal{M}$), an upper limit
\begin{equation}
\label{eq:ineq1}
    F \leq \mathcal{F} \equiv \sup_{\mathcal{M}} F
\end{equation}
for the Fisher information is obtained, where $ \mathcal{F}$ is commonly referred to as the \textit{quantum Fisher information} (QFI) and depends solely on the initial state $ \rho_0 $ and the dynamics encoding the parameter.

One can show\,\cite{BraunsteinCaves1994, BraunsteinCavesMilburn1996} that the QFI of a pure state $ \ket{\psi_0} $ under the unitary displacement\,\eqref{eq:App_Displ} is given by
\begin{equation}
\label{eq:QFI-unitary-appendix} \mathcal{F}_{\ket{\psi_0}} = 4\, \mathrm{Var}_{\ket{\psi_0}}(\hat{x}) \, .
\end{equation}
Furthermore the projective measurement $ \mathcal{M}=\lbrace \vert \psi_0 \rangle \langle \psi_0 \vert, \mathds{1}-\vert \psi_0 \rangle \langle \psi_0 \vert \rbrace $ onto the initial state actually optimizes the QFI in the limit of vanishing displacement for arbitrary initial states.
This is seen by expanding the probability $ P $ in orders of the displacement $ \oneddrift \ttime $:
\begin{equation}
P = \big\vert \bra{\psi_0} e^{i \oneddrift \ttime \hat{x}} \ket{\psi_0} \big\vert^2 = 1 - (\oneddrift \ttime)^2 \big( \bra{\psi_0} \hat{x}^2 \ket{\psi_0} - \bra{\psi_0} \hat{x} \ket{\psi_0}^2 \big) + \mathcal{O}\big((\oneddrift \ttime)^4\big)
\end{equation}
which allows to evaluate the FI of the binary measurement for displacements $\oneddrift \ttime$ as
\begin{equation}
\label{eq:FI-appendix} F = \dfrac{1}{P (1-P)} \left(\dfrac{\partial P}{\partial (\oneddrift \ttime)}\right)^2 \, .
\end{equation}
This shows the ideal character of the measurement because
\begin{equation}
\lim_{\oneddrift \ttime \to 0} F = 4  \left( \bra{\psi_0} \hat{x}^2 \ket{\psi_0} - \bra{\psi_0} \hat{x} \ket{\psi_0}^2 \right) = \mathcal{F}_{\ket{\psi_0}} \, .
\end{equation}
But it should be noted that this is only the case because, when calculating the Fisher information, numerator and denominator have the same dependence on $ \oneddrift \ttime $ to lowest order.

With this, we evaluate the QFI for momentum squeezed states $ \ket{\psi_0} = \ket{\xi}=e^{-r( a^2- a^{\dagger \, 2})/2} \ket{0}$, Fock states $\ket{n}$ and Schr\"{o}dinger-cat states $ \ket{\psi_0}= \ket{cat} =\mathcal{N}( \ket{\beta} + \ket{-\beta})$ with separation $2\beta$ between the coherent states:
\begin{align}
\mathcal{F}_{\ket{\xi}} &= 4\, \mathrm{Var}_{\ket{\xi}}(\hat{x}) = 2 e^{2r}  	\\
\mathcal{F}_{\ket{n}} &= 4\, \mathrm{Var}_{\ket{n}}(\hat{x}) = 2(2 n + 1)		\\
\mathcal{F}_{cat} &= 4\, \mathrm{Var}_{\ket{cat}}(\hat{x}) = 2 (1 + 8 \beta^2 \,\mathcal{N}^2)\,.
\end{align}
All these states asymptotically scale as $ \mathcal{F} \propto \bar{n} $ for $ \bra{\psi_0} a^{\dagger} a \ket{\psi_0} = \bar{n} \gg 1$.
Another relevant result for the study in the main text is that of a linear combination of Fock states $\ket{\psi_0}= \ket{\psi_{2+4}} = c_2 \ket{2} + c_4 \ket{4} $ can be optimized in terms of the QFI.
In this case, the QFI is
\begin{align}
\mathcal{F}_{2+4} 	&= 4 \left( \bra{\psi_{2+4}} \hat{x}^2 \ket{\psi_{2+4}} - \bra{\psi_{2+4}} \hat{x} \ket{\psi_{2+4}}^2 \right) \nonumber \\
\label{eq:DD-QFI_Displacement_2and4}	&= 2 \left( 1 + 4\, c_2^2 + 2 \sqrt{12}\, c_2 c_4 + 8\, c_4^2 \right)
\end{align}
with a maximum $ \mathcal{F}_{2+4} = 22 $ for $ c_2 = 1/2 $, $ c_4 = \sqrt{3}/2 $. A state close to this one is also obtained for the numerical optimization with small diffusion discussed in the main text.

As claimed in the main text, we will now show that any loss of contrast in the measurement breaks the QFI optimality at $ \oneddrift \ttime \to 0 $.
To see this, consider an imperfect POVM
\begin{equation}
\mathcal{M}_{\eta} =  \bigg\lbrace M^{(0)}_{\eta}, \, M^{(1)}_{\eta} = \mathds{1}-M^{(0)}_{\eta} \bigg\rbrace
\end{equation}
with $ 0 \leq \eta \leq 1/2 $ and
\begin{equation}
M^{(0)}_{\eta} = \big(1-\eta\big) \ket{\psi_0} \bra{\psi_0} + \eta \bigg(\mathds{1}- \ket{\psi_0} \bra{\psi_0}\bigg)
\end{equation}
for which $ (1-\eta)$ and $ \eta $ can be identified as a success rate and dark count rate of the measurement, respectively and the POVM considered in the main text is retrieved with $ \eta = 0 $.
These imperfections cause a loss of contrast as the measured probability can be expressed by
\begin{equation}
\tilde{P} = \big( 1-2 \eta \big) P + \eta
\end{equation}
in terms of the ideal $P$ from $ \mathcal{M} $.
The FI for this faulty POVM is then
\begin{equation}
\tilde{F} = \dfrac{ \left(\dfrac{\partial P}{\partial (\oneddrift \ttime)}\right)^2 }{P \left( 1 - P \right)- \dfrac{\eta (1-\eta)}{\big(1-2 \eta \big)^2}}
\end{equation}
in contrast to Eq.\,\eqref{eq:FI-appendix}.
This FI is reduced compared to the ideal case for any $ \eta > 0 $.
Moreover, due to the additional terms, the denominator will not vanish in the limit $ \oneddrift \ttime \to 0 $\,($ P \to 1 $) but the numerator still does, which reduces the FI to zero at this point even for a small loss of contrast and makes this point unfit for realistic experiments.
The dependence of FI on $ \tilde{P} $ for different losses $\eta$ is displayed in Fig.\,\ref{fig:reducedEff} for the ground state and a cat state with $ \bar{n} = 4 $.

\begin{figure*}[tb]
\centering
\includegraphics[width=\textwidth]{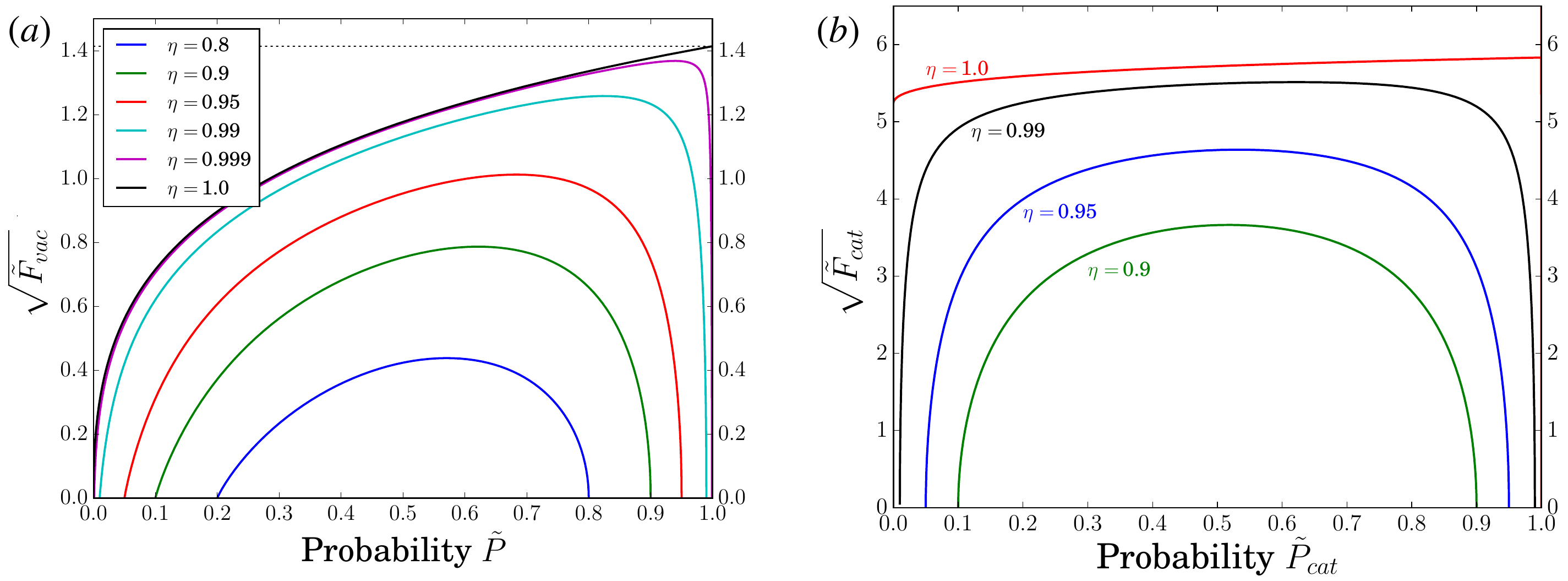}
\caption{Reduction of FI for unitary displacement with inefficient state detection. (a) Ground state $ \sqrt{\tilde{F}_{vac}} $ as a function of probability $ \tilde{P} $ for various detection efficiency $\eta$ (the curves are for decreasing values of $\eta$ from top to bottom). The dotted line at the top is $\sqrt{2}$, the QFI for the ground state. (b) FI $ \sqrt{\tilde{F}_{cat}} $ on the first oscillation of $ \tilde{P}_{cat} $.}
\label{fig:reducedEff}
\end{figure*}

\section{Influence of phase mismatch for squeezed states}\label{subsec:mismatch}

Realistic implementations of PRS protocols come with many additional restrictions.
Here the particular case of measurement inaccuracy and the ensuing impact on $\vert S \vert$ for squeezed states is discussed.

The recoil sensitivity of a squeezed state as shown in e.\,g. Fig.\,\ref{fig:DD-IndependentDiffVgl}a or \ref{fig:squeezed-states}a assumes ideal preparation and measurement of the motional state.
In particular, both squeezing phases $\phi$ and $\phi'$ for the initial state $\ket{\xi=r e^{i\phi}}$ and the projection state in the measurement
\begin{equation}
\mathcal{M}_{\ket{\xi'}} = \left\lbrace \ket{\xi'=r e^{i\phi'}}\bra{\xi'=r e^{i\phi'}},\, \mathds{1}-\ket{\xi'}\bra{\xi'} \right\rbrace
\end{equation}
should match exactly, i.e.\,$\phi=\phi'=\pi/2$ for squeezing in $p$ direction.
Any deviation of $ \Delta\phi = \phi-\phi' $ from zero leads to a slower decrease of the overlap under displacement, as well as a reduced initial overlap for larger detunings and therefore causes a loss in sensitivity.
The effect of such a mismatch upon states of different squeezing strength $r$ and for different ratios $ \oneddiff/\oneddrift $ is shown in Fig.\,\ref{fig:phase-mismatch}.
From part\,(a) one can infer that the width of acceptable $\Delta\phi$ changes only weakly for different ratios $ \epsilon $.
Only for larger noise does the distribution narrow down, with a reduced peak sensitivity.
The squeezing strength for the calculations in Fig. \ref{fig:phase-mismatch}a is $12.5$\,dB, so that a vertical section along $\Delta\phi=0$ corresponds exactly to the blue dashed curve in Fig.\,\ref{fig:DD-IndependentDiffVgl}a.
In contrast to the dependence on $\epsilon$, the sensitivity $\vert S \vert $ varies considerably with respect to the squeezing parameter $r$, as seen in part\,(b) of Fig.\,\ref{fig:phase-mismatch}.
For stronger squeezing the sensitivity becomes narrowly peaked around zero phase mismatch and decreases even below the vacuum level for larger deviations.
This means that, although more squeezing strength leads to an improved recoil sensitivity, the steep decline for phase mismatches might pose a practical limitation on the range of usable squeezing parameters due to the increased requirements on state preparation and detection.

\begin{figure*}[tb]
\centering
\includegraphics[width=\textwidth]{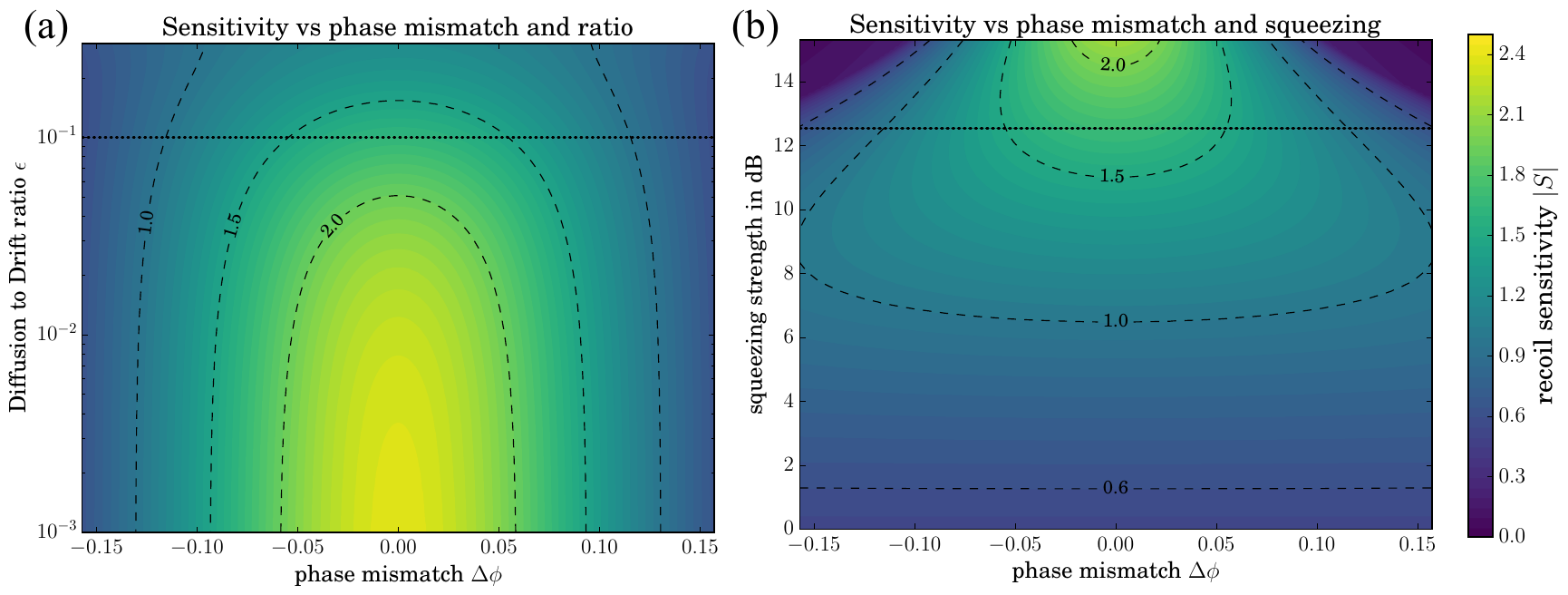}
\caption{Recoil sensitivity for squeezed states with a phase mismatch $\Delta\phi$ (in rad), compared to (a)\,the ratio $\epsilon = \oneddiff/\oneddrift $ and (b)\,the squeezing strength $r$ in dB. The displayed dotted lines are identical sections within both diagrams: In (a) this corresponds to a squeezing of $12.5$\,dB used for all ratios and in (b) to the ratio $\epsilon=0.1$ used for all values of $r$.}
\label{fig:phase-mismatch}
\end{figure*}

\section{Numerical methods}\label{sec:app-numerics}

Optimization over the initial Fock state combinations were performed using the \textit{L-BFGS-B} method, as implemented in SciPy\,\cite{SciPy}.
The \textit{L-BFGS-B} algorithm\,\cite{ByrdLuNocedalEtAl1995, ZhuByrdLuEtAl1997} is used for bound constrained minimization and is based on the quasi-Newton method of Broyden, Fletcher, Goldfarb, and Shanno (\textit{BFGS})\,\cite{NocedalWright2006}\,(pp.\,136). The bound is used here to optimize over the set of all properly normalized combinations of Fock states. The \textit{L-BFGS-B} algorithm showed the best stability for the optimization of Fock state superpositions.
All sensitivity calculations for the considered states showed good agreements to the solutions of an equivalent drift and diffusion master equation simulated with the QuTiP library\,\citep{JohanssonNationNori2013,QuTiP}.

\section{Average absorbed photon number}\label{sec:app-average}

Here we review a method to calculate the average number of absorbed photons based on the solutions of the optical Bloch equations. This follows the derivation of Cohen-Tannoudji et\,al.\,given in\,\cite{Cohen-Tannoudji_APINT_04}.
Ultimately, a mean number of absorbed photons per unit time will be obtained from dividing the mean absorbed power by the photon energy and then integrating over the pulse duration to find the average number of absorbed photons.\\
At first, one finds that the driving electric field $ \mathcal{E}_0\, \mathrm{cos}(\wlaser t) $ moves the electron from an initial position $r$ to $r + \mathrm{d}r$ in a time step from $t$ to $ t + \mathrm{d}t $. This process carries out the work
\begin{equation}
\mathrm{d}W = e \mathcal{E}_0\, \mathrm{cos}(\wlaser t) \cdot  \mathrm{d}r
\end{equation}
on the electron with charge $e$.
The average power for the absorption is then
\begin{equation}
\label{eq:Power} \left\langle\dfrac{\mathrm{d}W}{\mathrm{d}t}\right\rangle = \mathcal{E}_0\, \mathrm{cos}(\wlaser t)  \cdot \langle \dot{\mathrm{d}} \rangle
\end{equation}
with average dipole moment $ \langle \mathrm{d} \rangle = e \langle r \rangle $.
The expectation values therein can be expressed through $ \langle \sx \rangle $ and $ \langle \sy \rangle $, which are solutions of the optical Bloch equations in a rotating frame. This gives
\begin{equation}
\langle \mathrm{d} \rangle = \mathrm{d}_{dip} \cdot \bigg( \langle \sx \rangle\, \mathrm{cos}(\wlaser t) - \langle \sy \rangle\, \mathrm{sin}(\wlaser t) \bigg) \, .
\end{equation}
Inserting this into Eq.\,\eqref{eq:Power} and averaging over an oscillation period of the fast optical frequency $ \wlaser $ gives
\begin{align}
\overline{\left\langle\dfrac{\mathrm{d}W}{\mathrm{d}t}\right\rangle} &= - \mathrm{d}_{dip} \cdot \mathcal{E}_0 \wlaser \big[ \overline{\mathrm{cos}^2(\wlaser t)}\,\, \langle \sy \rangle + \\ &\hspace{1.5cm} \overline{\mathrm{sin}(\wlaser t)\,\, \mathrm{cos}(\wlaser t)}\,\, \langle \sx \rangle \big] \nonumber \\
 &= \dfrac{\Omega}{2} \wlaser \langle \sy \rangle
\end{align}
with the Rabi frequency $ \Omega = - \mathrm{d}_{dip} \cdot \mathcal{E}_0 $.
Finally, dividing by the energy $ \wlaser $ of each incident photon gives the mean number of photons per unit time and this can be integrated to find the mean number of absorbed photons per single laser pulse:
\begin{equation}
\label{eq:AveragePhotons} \langle N_1 \rangle = \dfrac{\Omega}{2} \int_0^{\tpulse} \mathrm{d}t\, \langle \sy (t) \rangle
\end{equation}
This coincides with a more intuitive calculation for PRS in which a single recoil event transfers momentum $\sqrt{2}\LD$ and the displacement $ \oneddrift \ttime = \Nph \sqrt{2}\LD$ is assumed to be obtained by $\Nph$ absorbed photons\,\footnote{This is up to the geometric factor $ \mathrm{cos}(\wmode t)$ in the solution for $ \oneddrift $ that expresses the contribution of photon absorption to a change in position or momentum}.
For $ \ttime $ pulses Eq.\,\eqref{eq:AveragePhotons} yields then
\begin{equation}
\Nph = \ttime \langle N_1 \rangle
\end{equation}
absorbed photons, or rearranged in terms of $ \ttime $ this gives the interaction time $\tspec$ necessary for a single absorption on average.

\end{widetext}
\end{appendix}

\end{document}